\documentclass[
    aps,
    pra,
    amsmath,amssymb,
    preprint, single column
    superscriptaddress
    ]{revtex4-1}

\usepackage{graphicx}
\usepackage{subcaption}
\usepackage{latexsym}
\usepackage{physics}
\usepackage{amsmath}
\usepackage{color}
\usepackage[normalem]{ulem}

\usepackage{float}
\usepackage[varg]{txfonts}
\usepackage[svgnames]{xcolor}
\usepackage[
     colorlinks        = true,
     unicode           = true,
     pdfstartview      = FitV,
     linktocpage       = true,
     linkcolor         = OrangeRed,
     citecolor         = MediumSeaGreen,
     urlcolor          = RoyalBlue,
     bookmarks         = true,
     bookmarksnumbered = true,
     breaklinks=true,
     pdftitle={},
     pdfauthor={}
]{hyperref}

\begin{document}

\title{Fixing Divergence in Carleman Linearization via Analytical Continuation}

\author{Mingshuo Zhu}
\email{mingshuo@qunasys.com}
\affiliation{QunaSys Inc., 1-13-7 Hakusan, Bunkyo, Tokyo 113-0001, Japan}

\author{Hayato Higuchi}
\email{higuchi@qunasys.com}
\affiliation{QunaSys Inc., 1-13-7 Hakusan, Bunkyo, Tokyo 113-0001, Japan}

\author{Hokuto Iwakiri}
\email{iwakiri@qunasys.com}
\affiliation{QunaSys Inc., 1-13-7 Hakusan, Bunkyo, Tokyo 113-0001, Japan}

\author{Kouki Nakamura}
\email{nakamura@qunasys.com}
\affiliation{QunaSys Inc., 1-13-7 Hakusan, Bunkyo, Tokyo 113-0001, Japan}

\author{Naohisa Sueishi}
\email{sueishi@qunasys.com}
\affiliation{QunaSys Inc., 1-13-7 Hakusan, Bunkyo, Tokyo 113-0001, Japan}

\author{Shih-Yen Tseng}
\email{tseng@qunasys.com}
\affiliation{QunaSys Inc., 1-13-7 Hakusan, Bunkyo, Tokyo 113-0001, Japan}

\author{Shoichiro Tsutsui}
\email{tsutsui@qunasys.com}
\affiliation{QunaSys Inc., 1-13-7 Hakusan, Bunkyo, Tokyo 113-0001, Japan}

\date{\today}

\begin{abstract}
Nonlinear differential equations play a crucial role in modeling a wide range of phenomena,
yet their solutions remain notoriously difficult to obtain. 
With the rapid development of quantum computing, quantum algorithms for efficiently solving such equations are actively being explored. 
One promising approach is based on Carleman linearization, which transforms nonlinear differential equations into linear systems. However, 
this method suffers from exponential divergence beyond a certain time scale. 
By reformulating the solutions in terms of eigenvalues and eigenvectors, 
we identify that this divergence originates from the Laurent expansion outside its neighborhood of convergence. 
To address this issue, we insert a regularized function to the divergent solution hinted by analytical continuation.
We validate this divergence-correction method on both the logistic equation and some other partial differential equations like KPP-Fisher equations and Phase-Field models under periodic conditions. 
We implement our method for the logistic equation using the Linear Combination of Unitaries (LCU) quantum algorithm, providing a detailed complexity and error analysis.
\end{abstract}

\maketitle

\section{Introduction}

Nonlinear differential equations play a pivotal role in modern mathematical modeling \cite{TMS2nd}. They govern fluid transport and turbulence (Navier–Stokes equations), pattern formation and front propagation in chemistry and biology (reaction–diffusion systems such as Kolmogorov–Petrovsky–Piskunov–Fisher (KPP–Fisher) \cite{Kolmogorov1937, Fisher1937}), order-parameter dynamics in materials science (Ginzburg–Landau \cite{Landau1937, GinzburgLandau1950} and phase–field models \cite{CahnHilliard1958, AllenCahn1979, KarmaRappel1996}), wave propagation and optimal control and reinforcement learning through Hamilton–Jacobi–Bellman (HJB) equations \cite{Bellman1957}). Across these domains, nonlinearity introduces multiscale coupling, bifurcations, and potential blow-up, all of which challenge classical discretizations that must balance accuracy, stability, and rapidly growing computational costs—especially in high dimensions.

Against this backdrop, huge progress has been made in solving linear differential equations on a quantum computer. When the system is unitary, quantum algorithms promise polynomial—or in some settings exponential—speedups for linear algebraic primitives that underlie many discretizations of differential equations \cite{Babbush:2023uqz}. Several approaches have been developed to unitarize and solve non-unitary linear systems of differential equations. In general, Linear-Combination-of-Unitaries (LCU) \cite{Childs:2019hts, Meister:2020qfp} and Quantum Singular Value Transformation (QSVT) \cite{Gilyen:2018khw, Martyn:2021eaf} provide a mature toolkit for simulating sparse linear dynamics and solving linear systems.
In particular, Linear Combination of Hamiltonian Simulation (LCHS), developed within the LCU framework, represents non-unitary dynamics as a continuous linear combination of unitary Hamiltonian evolutions \cite{An:2023tzn,PhysRevApplied.23.014063}.
More specifically, the Black-Scholes equation in Finance can be solved on a quantum computer by constructing an extended system of non-unitary systems to be unitary \cite{Gonzalez-Conde:2021zbv}. 
The diffusion equation can also be treated by Schrödingerisation, which introduces an auxiliary spatial dimension to map non-unitary dynamics to unitary Schrödinger evolution and is closely related to the LCHS framework \cite{Jin:2022nwz}.

Extending these advantages to nonlinear dynamics, however, is nontrivial and still in the primitive stage of research, because the quantum evolution is linear, and one needs to first transform nonlinear systems into linear ones. Several quantum strategies have emerged \cite{Lloyd:2020xda}, notably, Carleman linearization, which lifts a nonlinear system into an infinite-dimensional linear system and then truncates part of the array, and the algorithm based on it has been proposed recently for dissipative nonlinear differential equations~\cite{Liu:2020dof,an2023efficientquantumalgorithmnonlinear,Krovi2023improvedquantum,Costa_2025,jennings2025quantumalgorithmsgeneralnonlinear}. However, it has been pointed out that the Carleman linearization for general cases may suffer from exponential divergence beyond a certain time limit. Even in the widely used lattice Boltzmann method (LBM) \cite{fluids7010024,Itani:2023ltd,Sanavio:2024tnw}, a flagship approach in solving Navier–Stokes equations, the Carleman solutions diverge after a finite time; the same behavior is observed for the logistic, KPP–Fisher, and phase–field models in \cite{Liu:2022eio, Endo:2024wrt}, where divergence is either avoided by restricting the parameters of the differential equations or mitigated by shifting the expansion point along the time evolution. A similar method by partitioning space into multiple regions where the center and size of the embedding region are chosen to control convergence is proposed to simulate the chaotic dynamical systems with Carleman linearization \cite{Novikau:2025zfr}.

Motivated by these observations, we turn to an analysis of the origin of this divergence. By reformulating the solutions in terms of eigenvalues and eigenvectors, we identify that the observed divergence arises from evaluating a Laurent expansion of the solution based on the Carleman linearization outside its radius of convergence. In particular, we perform a detailed analysis of the divergence phenomenon in simulations of the logistic equation and compare the resulting behavior with the exact solution. Our findings agree with the observations in~\cite{Endo:2024wrt}, where the divergence is mitigated by shifting the expansion point along the time evolution. 

To remedy this issue, we propose an analytical–continuation approach that enlarges the region of convergence. A suitable conformal mapping, chosen to avoid the influence of nearby singularities, guarantees convergence along the entire positive real axis. Furthermore, we find this divergence‑correction method can be simply realizing by inserting a regularized function into the divergent Laurent expansion, and validate this regularized Carleman method on the three non-dissipative types of diffusion-reaction differential equations including the logistic equation, the KPP–Fisher equation and the phase–field model. Especially for the phase-field model, the cubic nonlinearity necessitates a distinct conformal map.
In all cases, we also identify a constraint relating the initial state and the scale parameter of the conformal map which determines whether convergence can be achieved. 

Finally, we present a quantum implementation of the regularized Carleman method for the logistic equation. The procedure is as follows. Since the Carleman matrix is generally non-unitary and spare, we employ block encoding \cite{Camps:2022jnx} to embed it into a unitary matrix. Next, we construct an eigenvalue transformation that maps the eigenvalues of the Carleman matrix to those of the regularized function, and implement this transformation on the Carleman matrix via the LCU method, which requires expanding the regularized function in terms of monomial polynomials. Finally, by applying the resulting operator to the initial state and measuring the first component, we obtain the regularized Carleman solution without divergence. 

This paper is organized as follows. Section~\ref{sec:rev_CL} reviews Carleman linearization and introduces notation for the lifted variables and the resulting infinite block system. Especially, Section~\ref{sec:divergence} identifies the origin of exponential divergence through a binomial--eigenbasis representation and Section~\ref{sec:ana_cont} presents the analytic-continuation method based on a Möbius conformal map, derives an explicit convergence condition and construct the regularized function to fix the divergence. Furthermore, it demonstrates the approach on the logistic equation in numerical calcualtion. Section~\ref{sec:KPP} applies the same strategy to KPP-Fisher equations with periodic boundary conditions, including the $L=3$ (integer spectrum, Laurent structure) and $L=5$ (multi-series, non-integer spectrum) cases, together with associated stability properties. Section~\ref{sec:PFM} addresses divergence in the cubic logistic equation and phase–field models, and introduces a modified conformal mapping tailored to these systems. Section~\ref{sec:QSVT} gives a concrete quantum implementation of the regularized Carleman method and related resource estimation for the logistic equation. Finally, Section~\ref{sec:conclusion} summarizes our results, and discusses open problems, including the development of efficient methods for constructing block encodings of Carleman matrices for general classes of differential equations and the application of the present regularization strategy to computational fluid dynamics.

\section{Review of Carleman Linearization}\label{sec:rev_CL}

We briefly review Carleman linearization \cite{Carleman:1932}. Consider the autonomous nonlinear ODE
\begin{equation}
  \frac{d}{dt}\,x=f\bigl(x\bigr),\qquad x(0)=x_0\in\mathbb{R}^n,
  \label{eq:nonlinear-ode}
\end{equation}
where $f:\mathbb{R}^n\to\mathbb{R}^n$ is analytic in a neighborhood of $x_0$. While Carleman linearization applies to general analytic vector fields, its formalism is most transparent for polynomial systems. Many nonlinear systems can be recast in polynomial form by augmenting the state with auxiliary variables; in any case, an analytic $f$ admits a truncated Taylor approximation.

Without loss of generality, consider an $n$–dimensional system with polynomial nonlinearity of maximal degree $k$:
\begin{equation}
  \frac{d}{dt}\,x \;=\; F_1 x \;+\; F_2\, x^{[2]} \;+\; \cdots \;+\; F_k\, x^{[k]},
  \label{eq:poly-ode}
\end{equation}
where $F_j\in \mathbb{R}^{n\times n^j}$ are constant matrices and $x^{[j]}$ denotes the $j$th Kronecker (tensor) power of $x$:
\begin{equation}
  x^{[j]} \;:=\; \underbrace{x \otimes x \otimes \cdots \otimes x}_{j~\text{times}}
  \in \mathbb{R}^{n^j},
  \label{eq:kronecker-power}
\end{equation}
so that $x^{[2]}$ stacks all quadratic monomials in the components of $x$.

The core idea of Carleman linearization is to embed the original nonlinear ODE into a higher-dimensional linear system by introducing a hierarchy of lifted variables. Specifically, we define
\begin{equation}
  y_i := x^{[i]},\qquad i\in\mathbb{N}.
  \label{eq:yi-def}
\end{equation}
Differentiating $y_i$ and repeatedly applying the product rule yields linear relations between $\dot y_i$ and $\{y_j\}_{j\ge i-1}$. If we collect $\mathbf{y} \;=\; (y_1,y_2,\ldots)^\top$,
the Carleman lift takes the compact linear form
\begin{equation}
  \frac{d}{dt}\,\mathbf{y} \;=\; \mathbf{A}\,\mathbf{y},
  \label{eq:carleman-inf}
\end{equation}
where $\mathbf{A}$ is an infinite, block-structured matrix (upper block-Hessenberg for polynomial systems). For numerical computation, one truncates at level $K$ by projecting Eq.~\eqref{eq:carleman-inf} onto the finite subspace spanned by $\{y_1,\ldots,y_K\}$ and neglecting the coupling to $y_{K+1}$ and higher.

\subsection{The Logistic Equation}
The logistic equation is
\begin{equation}
  \frac{dx}{dt}=x(1-x),
\end{equation}
whose exact solution is
\begin{equation}\label{eq:exact_log}
  x(t) \;=\; \frac{x_0\,e^t}{1 + x_0\,(e^{t}-1)},
\end{equation}
which holds for any $x_0\in\mathbb{R}$. Unless otherwise stated, we take $x_0\in(0,1)$ to align with the Carleman–convergence analysis that follows.

Since the logistic equation is a scalar instance of Eq.~\eqref{eq:poly-ode}, its Carleman lift is particularly simple. Define the infinite vector
\[
  \mathbf{y}=(y_1,y_2,\ldots)^\top,\qquad
  y_1=x,\; y_2=x^2,\; y_3=x^3,\;\ldots.
\]
For $k\ge 1$, one obtains
\begin{equation}\label{eq:CL_decomp}
  \frac{d y_k}{dt} \;=\; k\bigl(y_k - y_{k+1}\bigr).
\end{equation}
Equivalently, the infinite matrix $\mathbf{A}$ has diagonal entries $A_{k,k}=k$ and super-diagonal entries $A_{k,k+1}=-k$ (with $k\ge 1$).

\begin{figure}[t]
  \centering
  \includegraphics[width=0.8\linewidth]{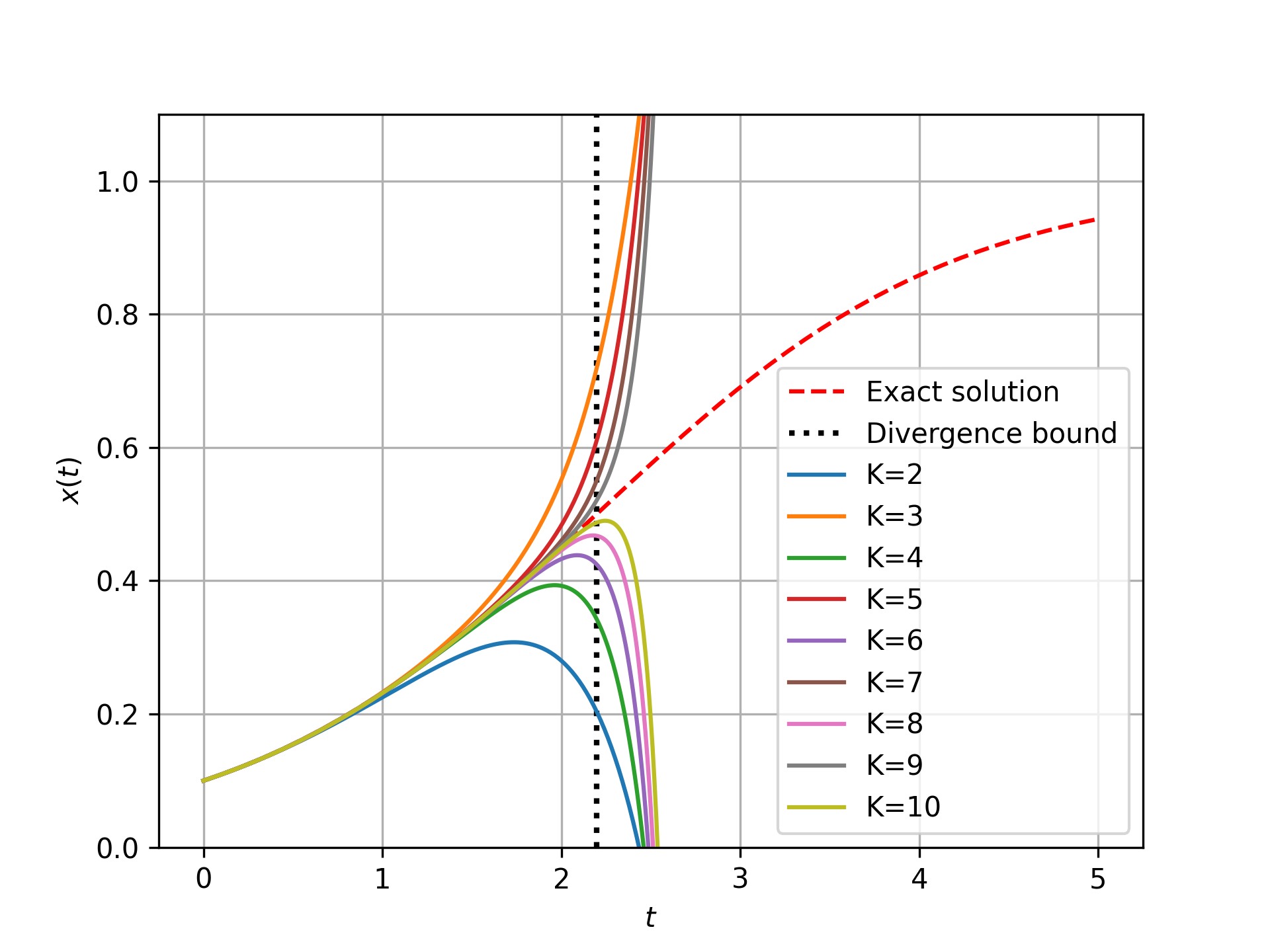}
  \caption{Comparison of Carleman linearization with the exact solution of the logistic equation, a prototype model for saturation in population dynamics, with $x_0=0.1$.  The dotted curve plots the exact solution \eqref{eq:exact_log}, while the solid curves show Carleman solutions with increasing truncation orders $K$.}
  \label{fig:logistic_cl}
\end{figure}

To compute numerically, fix a truncation order $K$ and solve the $K$–dimensional linear system obtained from Eq.~\eqref{eq:CL_decomp} by discarding the coupling to $y_{K+1}$ (i.e., setting it to zero in the closure). As illustrated in Fig.~\ref{fig:logistic_cl}, two salient features arise \cite{Itani:2021acl, Endo:2024wrt}:

\begin{itemize}
  \item For sufficiently large $K$, the truncated linear system approximates the nonlinear dynamics with high accuracy over a finite time interval.
  \item Regardless of how large $K$ is (and even in the formal infinite-dimensional limit), the Carleman series representation ultimately diverges at later times; in Sec.~\ref{sec:divergence} we prove this phenomenon and quantify its onset.
\end{itemize}
The divergence of the truncated series at long times motivates the analytic–continuation strategy developed in this work.

\subsection{Exponential Divergence}\label{sec:divergence}
Before turning to the fix strategy, let us trace the exponential divergence of the Carleman series to its finite radius of convergence, using the logistic equation as a testbed.
As in the review, Eq.~\eqref{eq:CL_decomp} induces the infinite-dimensional linear system \eqref{eq:carleman-inf} with solution
\begin{equation}\label{eq:gen_sol}
    \mathbf{y}(t) \;=\; \exp(\mathbf{A}t)\,\mathbf{y}(0).
\end{equation}
For the logistic case, the Carleman matrix $\mathbf{A}$ has a point spectrum
$\sigma(\mathbf{A})=\{\,\lambda_k=k:\;k\in\mathbb{Z}_+\,\}$ with a binomial
eigenbasis. In particular, one may choose the eigenvectors as
\begin{equation}\label{eq:logi_eigenv}
\mathbf{y}^{(1)}=\!\begin{pmatrix}1\\[2pt]0\\[2pt]0\\ \vdots\end{pmatrix},\quad
\mathbf{y}^{(2)}=\!\begin{pmatrix}-1\\[2pt]1\\[2pt]0\\ \vdots\end{pmatrix},\quad
\mathbf{y}^{(3)}=\!\begin{pmatrix}1\\[2pt]-2\\[2pt]1\\ \vdots\end{pmatrix},\quad
\mathbf{y}^{(4)}=\!\begin{pmatrix}-1\\[2pt]3\\[2pt]-3\\ \vdots\end{pmatrix},\;\ldots
\end{equation}
Especially, the first component is given by
\[
\bigl(\mathbf{y}^{(k)}\bigr)_1 = -(-1)^k.
\]
Expanding $\mathbf{y}(0)=\sum_{k\ge 1} a_k\,\mathbf{y}^{(k)}$ and inserting it into Eq.~\eqref{eq:gen_sol} yields
\begin{equation}\label{eq:carleman_solution}
    \mathbf{y}(t) 
    \;=\; \sum_{k=1}^{\infty} a_k\,\exp(kt)\,\mathbf{y}^{(k)}= \sum_{k=1}^{\infty} a_k\,\zeta^{k}\,\mathbf{y}^{(k)},
\end{equation}
where we set $\zeta := e^{t}$ for convenience. Matching the initial data for the logistic equation gives a closed form for the coefficients:
\begin{align}
    a_k
    &= \sum_{n=1}^{\infty} \frac{\Gamma(n+k-1)}{\Gamma(k)\,\Gamma(n)}\, x_0^{\,n+k-1}
      \;=\; \left(\frac{x_0}{1-x_0}\right)^{\!k},
      \quad \text{for } |x_0|<1. \label{eq:ck_closed}
\end{align}

From the first component (recall $y_1=x$) and Eq.~\eqref{eq:logi_eigenv} we obtain
\begin{align}
  x(\zeta)
  &= -\sum_{k=1}^{\infty} (-1)^k\, a_k\, \zeta^{k}
   \;=\; -\sum_{k=1}^{\infty}
       \left(\frac{x_0\,\zeta}{x_0-1}\right)^{\!k}
   \;=\; \frac{x_0 e^{t}}{1 - x_0 + x_0 e^{t}}, \label{eq:taylor_expansion}
\end{align}
which coincides with the exact solution \eqref{eq:exact_log} whenever the
geometric series converges, i.e.,
\[
\left|\frac{x_0 \zeta}{x_0-1}\right| < 1
\quad\Longleftrightarrow\quad
|\zeta| < \frac{1-x_0}{x_0}\,.
\]
For $x_0\in(0,1)$ this radius is finite, and since $\zeta=e^t$ increases without bound along the positive real axis, the Carleman–series representation necessarily becomes divergent at late times. This establishes the exponential (in $t$) divergence of the naive Carleman expansion even though the underlying solution remains analytic.

\section{Analytic Continuation for the logistic equation}\label{sec:ana_cont}
\begin{figure}[t]
  \centering
  \includegraphics[width=0.8\linewidth]{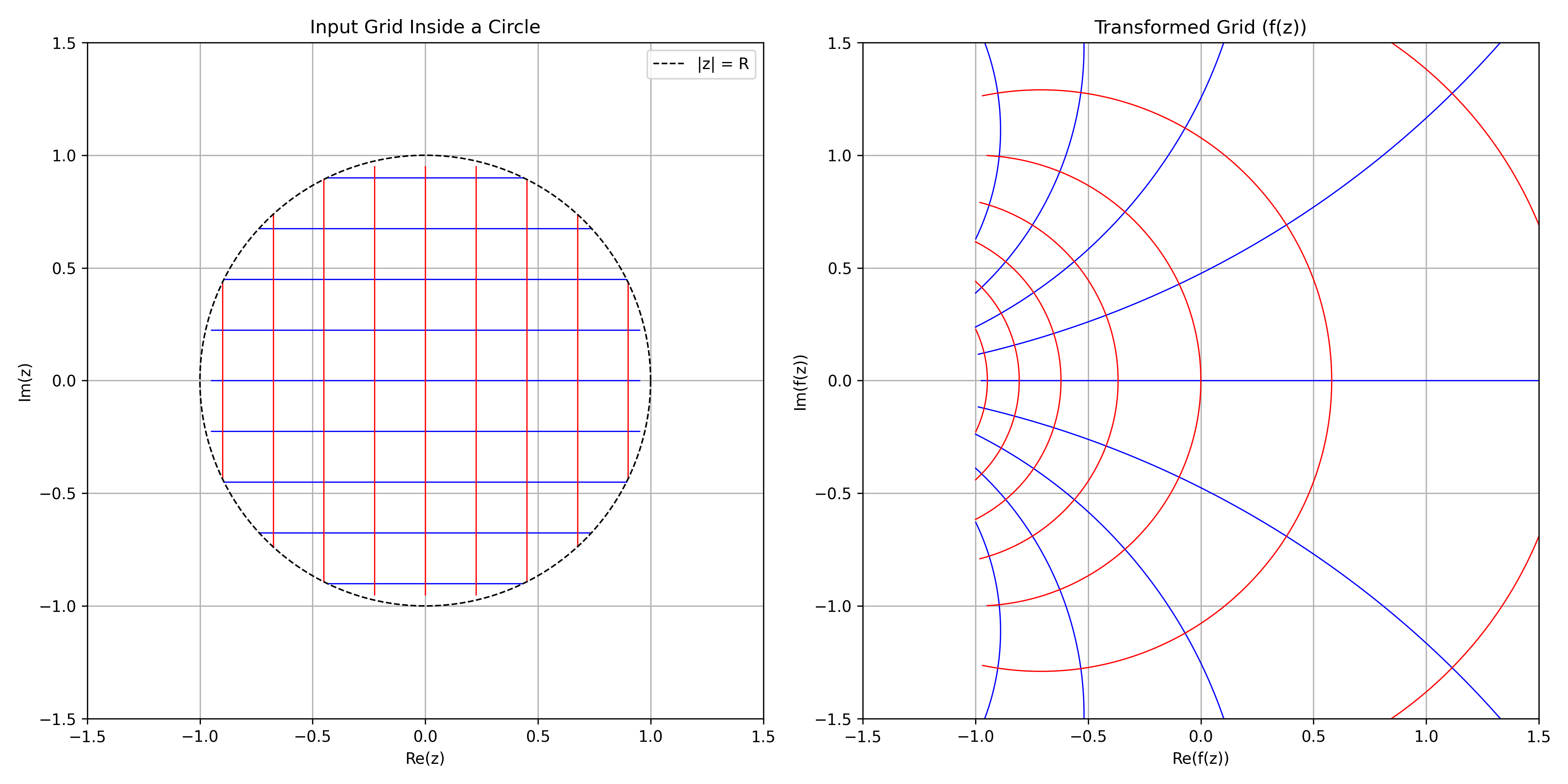}
  \caption{Conformal mapping between the $\omega$-plane and the $\zeta$-plane. The unit disc (left) represents the domain of $\omega$, while the region on the right represents the domain of $\zeta$.}
  \label{fig:conformal_map}
\end{figure}

The divergence above stems solely from the finite radius of convergence of the $\zeta$–series. A simple analytic continuation enlarges the effective domain. We adopt the Möbius (conformal) map as in Fig.\ref{fig:conformal_map}
\begin{equation}\label{eq:conformal_map}
     \zeta \;=\; \varphi(\omega) \;=\; \frac{c\,\omega}{1-\omega},
     \quad\leftrightarrow\quad
     \omega \;=\; \frac{\zeta}{\zeta + c},
\end{equation}
with real scaling parameter $c>0$. 
This choice has three practical properties followed by \cite{TakahasiMori:1973cat, Mori:1980, TakahasiMori:1984acs}:
\begin{itemize}
  \item It maps the positive real ray $\zeta\in[0,\infty)$ to $\omega\in[0,1)$, so time evolution ($t\ge 0$) stays inside the unit disk in the $\omega$–plane.
  \item $\varphi(0)=0$, ensuring that the coefficient of $\omega^m$ depends only on the coefficients up to order $m$ in the original series.
  \item The inverse is rational and trivial to evaluate, which simplifies both analysis and numerics.
\end{itemize}
The map \eqref{eq:conformal_map} is the special case $f(\omega)=\dfrac{c\omega+d}{a\omega+b}$ with $d=0$ and $b=1$ of a general Möbius transform.

We will still focus on the logistic equation first and generalize the conclusions in Sec.\ref{sec:KPP} and Sec.\ref{sec:PFM}.
Using the binomial expansion,
\begin{equation}\label{eq:binom_expansion}
\left(\frac{\omega}{1-\omega}\right)^{\!k}
= \sum_{n=1}^{\infty} \frac{\Gamma(n+k-1)}{\Gamma(k)\,\Gamma(n)}\,\omega^{\,n+k-1},
\end{equation}
and substituting $\zeta=\varphi(\omega)$ into Eq.\eqref{eq:taylor_expansion} with
$r:=\dfrac{c x_0}{1-x_0}$, we obtain after rearranging the sums
\begin{equation}\label{eq:logis_conmap}
    x(\omega)
    \;=\; r \sum_{n=1}^{\infty}
    \frac{\omega^{\,n}}{\alpha^{\,n}}
    \sum_{k=1}^{\infty}
    \frac{\Gamma(n+k-1)}{\Gamma(k)\,\Gamma(n)}\,\alpha^{\,n+k-1},
    \quad \alpha := -r\,\omega ,
\end{equation}
where $\Gamma(z)$ is the gamma function
\begin{equation}
\Gamma(z)=\int_0^{\infty} t^{z-1} e^{-t} d t, \quad \Re(z)>0
\end{equation}

Requiring absolute convergence for all $|\omega|<1$ is guaranteed by $|\alpha|<1$ on the unit disk, i.e.
\begin{equation}\label{eq:conver_cond}
    |r|<1
    \quad\Longleftrightarrow\quad
    c \;<\; \frac{1-x_0}{x_0}.
\end{equation}
Under Eq.\eqref{eq:conver_cond}, the inner sum in Eq.\eqref{eq:logis_conmap} is geometric and the series collapses to
\begin{equation}\label{eq:logis_simple}
    x(\omega)
    \;=\; r \sum_{n=1}^{\infty}
    \Bigl(\frac{\omega}{1-\alpha}\Bigr)^{\!n}
    \;=\; r \sum_{n=1}^{\infty}
    \left(\frac{\zeta}{(1+r)\,\zeta + c}\right)^{\!n},
\end{equation}
which converges for all $|\omega|<1$, hence for all $t\ge 0$ along the real time axis (since $\omega=\zeta/(\zeta+c)\in[0,1)$). In practice,  the outer sum in Eq.~\eqref{eq:logis_conmap} is truncated the outer sum at order $M$. Because $\varphi(0)=0$ (as noted below Eq.~\eqref{eq:conformal_map}), the coefficients of $\omega^n$ consist of finite terms $\{a_{i}, \; i \le n\}$, ensuring numerical tractability. The resulting approximation converges uniformly on compact subsets of the unit disk and remains well-behaved for arbitrarily large $t$, provided Eq.~\eqref{eq:conver_cond} is satisfied. The influence of $M$ on the convergence rate is analyzed in the following section.

\subsection{The regularized function: $f_{M}(k,t)$}
Finally, we assess the performance of the conformal–mapping approach in practical computations. Let us first turn back to Eq.\eqref{eq:carleman_solution}. In general, there is no explicit expression for the coefficients $\{a_k\}$, but we can still apply the conformal map and use the binomial expansion:
\begin{equation}\label{eq:numerical_carleman_solution}
    \mathbf{y}(t) 
    \;=\;  \sum_{k=1}^{\infty} a_k\mathbf{y}^{(k)}\zeta^{k} = \sum_{k=1}^{\infty} a_k\mathbf{y}^{(k)}\,\sum_{n=1}^{\infty} \frac{\Gamma(n+k-1)}{\Gamma(k)\,\Gamma(n)}\,c^k\omega^{\,n+k-1}
\end{equation}
Throughout, the Carleman linearization is truncated at order \(K\) (see Sec.~\ref{sec:rev_CL}). The solution \eqref{eq:numerical_carleman_solution} contains an infinite sum over \(n\), which should also be truncated for numerical evaluation. For convenience, let us first set $n+k-1 =m$, then the expression \eqref{eq:numerical_carleman_solution} becomes
\begin{equation}\label{eq:regularized_sol}
\begin{aligned}
    \mathbf{y}(t) 
    &= \sum_{k=1}^{\infty} a_k\mathbf{y}^{(k)}\,\sum_{m=1}^{M} \frac{\Gamma(m)}{\Gamma(k)\,\Gamma(m-k+1)}\,c^k\omega^{m} \\
    &= \sum_{k=1}^{\infty} a_k\zeta^k\mathbf{y}^{(k)}\; f_{M,c}(k,t)
\end{aligned}
\end{equation}
with the regularized function
\begin{equation}\label{eq:regularized_func}
    f_{M,c}(k,t) = \sum_{m=1}^{M} \frac{\Gamma(m)}{\Gamma(k)\,\Gamma(m-k+1)}\,c^k\omega^{m}\zeta^{-k}\in [0,1]
\end{equation}
Here the summation of $m$ begins from $1$ instead of $k$ because $\Gamma(m-k+1)$ hits a pole when $m<k$. If the regularized function reduce to 1, the solution \eqref{eq:regularized_sol} becomes nothing but the divergent one \eqref{eq:carleman_solution}. The regularized function $f_{M,c}$ approaches to $0$ as $t\rightarrow\infty$ to fix the divergence, especially in the reaction–diffusion equations, it becomes the regularized incomplete beta function
\begin{equation}
    f_{M,c}(k,t) = I_{1-\omega}(k, M-k+1)
\end{equation}

\begin{figure}[h]
  \centering
  \includegraphics[width=\linewidth]{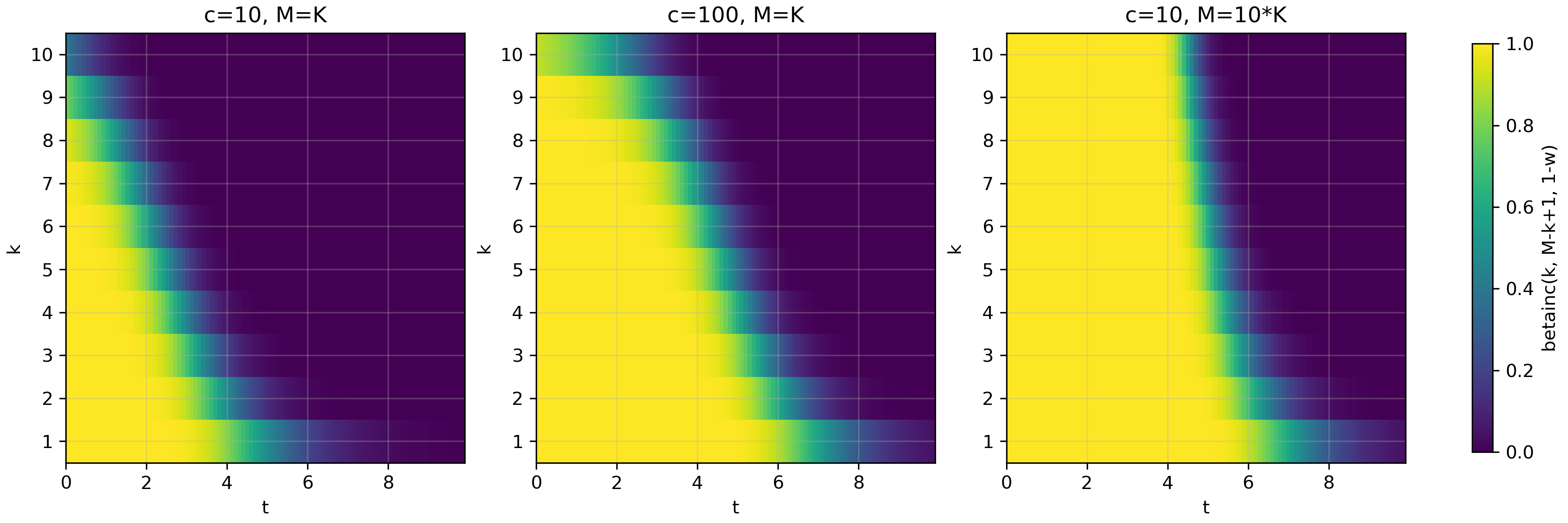}
  \caption{Comparison of heatmaps for the regularized incomplete beta function $\mathrm{I}_{1-w(t)}(k, M-k+1)$ with $w(t)=e^t/(e^t+c)$. The three panels correspond to, from left to right, $(c,M)=(10,K)$, $(100,K)$, and $(10,10K)$.}
  \label{fig:beta_fun}
\end{figure}

As in Figure~\ref{fig:beta_fun}, increasing $c$ shifts the transition in $t$, while increasing $M$ sharpens the dependence on the index $k$. 

\subsection{The Numerical Results}
\begin{figure*}[t]
    \centering
    \includegraphics[width=1.0\linewidth]{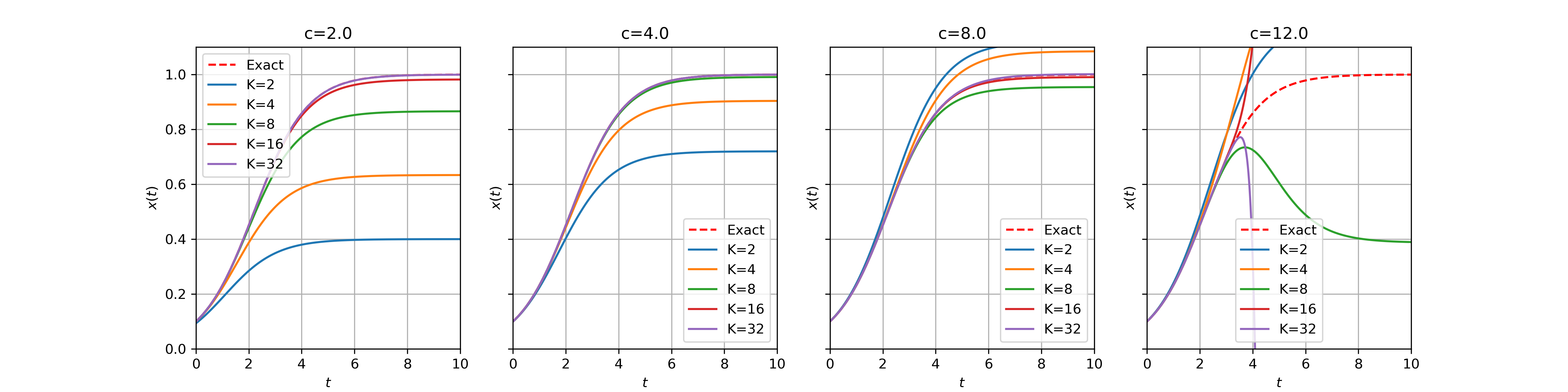}
    \caption{Time evolution of the logistic solution after conformal mapping for $x_0=0.1$ and several scale parameters $c$. Cases $c=2$ and $c=4$ exhibit stable convergence; $c=8$ shows non-monotone (unstable) convergence; and $c=12$ diverges, consistent with the analytic bound $c<(1-x_0)/x_0=9$.}
    \label{fig:Logis_conver}
\end{figure*}

Furthermore, let us also see how the parameters $\{M, c\}$ affects the final approximate solution. We fix the other parameters as \(t_0=0\), \(t_f=10\), time step \(\Delta t = 0.01\), and keep $M=K$ and vary $c$.  Figure~\ref{fig:Logis_conver} presents numerical results for several values of the scale parameter \(c\), illustrating how \(c\) influences convergence behavior and accuracy.

According to the constraint given in Eq.~\eqref{eq:conver_cond}, the relationship between $c$ and $x_0$ determines the convergence of the time evolution. For instance, when $x_0 = 0.1$, the conformal map yields convergence only if $c \leq 9$. As illustrated in Fig.~\ref{fig:Logis_conver}, the value of $c$ significantly affects the convergence behavior. For small values such as $c=2$, a large truncation parameter $K$ is required to approximate the exact solution. As $c$ increases, the convergence improves and becomes faster. For example, when $c=4$, the approximate curve coincides with the exact solution already at $K=8$, which is substantially more efficient than in the case of $c=2$. However, as $c$ approaches its limiting value (e.g., $c=8$), the convergence behavior deteriorates: although large-$K$ approximations initially improve the results, the convergence becomes less stable. Here, ``oscillatory’’ refers to the fact that smaller $K$ values can yield artificially higher convergent values. Finally, once $c$ exceeds its critical limit, the conformal mapping approach ceases to be valid.

Then, let us fix $c$ and vary $M$. The Carleman order $K$ determines how many lifted modes are retained, while $M$ determines how far the conformally mapped series is truncated. In theory, $M$ should approach infinity, suggesting that a larger $M$ can accelerate the convergence. However, in numerical computations, $M>K$ results in additional error and may cause the simulation to diverge. A more appropriate choice is therefore to set $M$ proportional to~$K$.

\begin{figure}[H]
    \centering
    \includegraphics[width=1.0\linewidth]{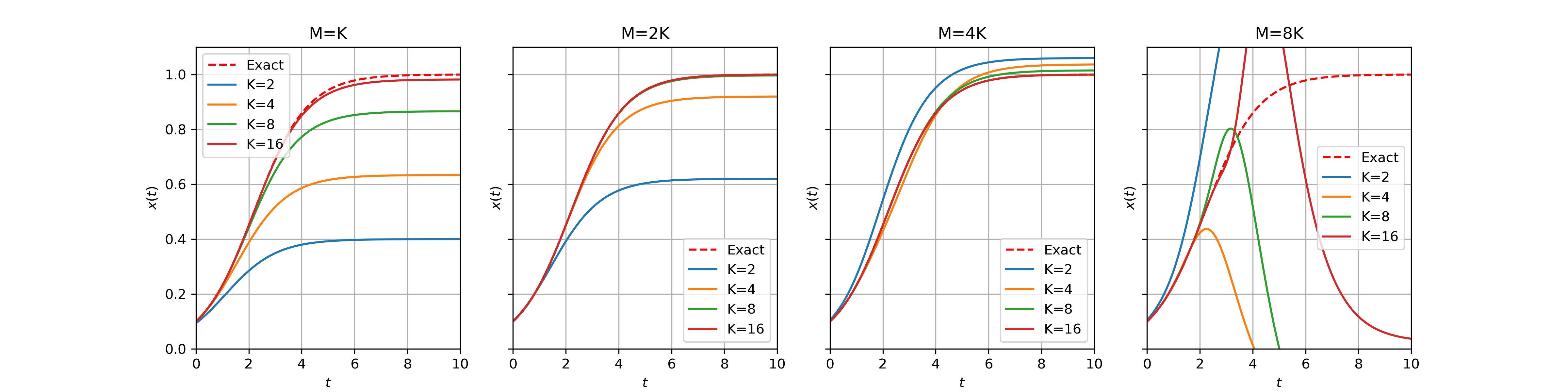}
    \caption{The Carleman solution to the logistic equation with different $M$, $x_0=0.1$, $c=2$ and truncation parameters up to $K=16$.}
    \label{fig:Logis_conver_M}
\end{figure}

Fig.\ref{fig:Logis_conver_M} presents simulation results for $M = K, 2K, 4K, 8K$ with initial value $x_0 = 0.1$ and $c = 2$. As shown, the convergence rate increases as $M$ becomes larger, but the simulation may become oscillatory or diverge when $M$ is excessively large relative to $K$ (e.g., the case $M = 4K$).

\begin{figure}[H]
    \centering
    \includegraphics[width=0.60\linewidth]{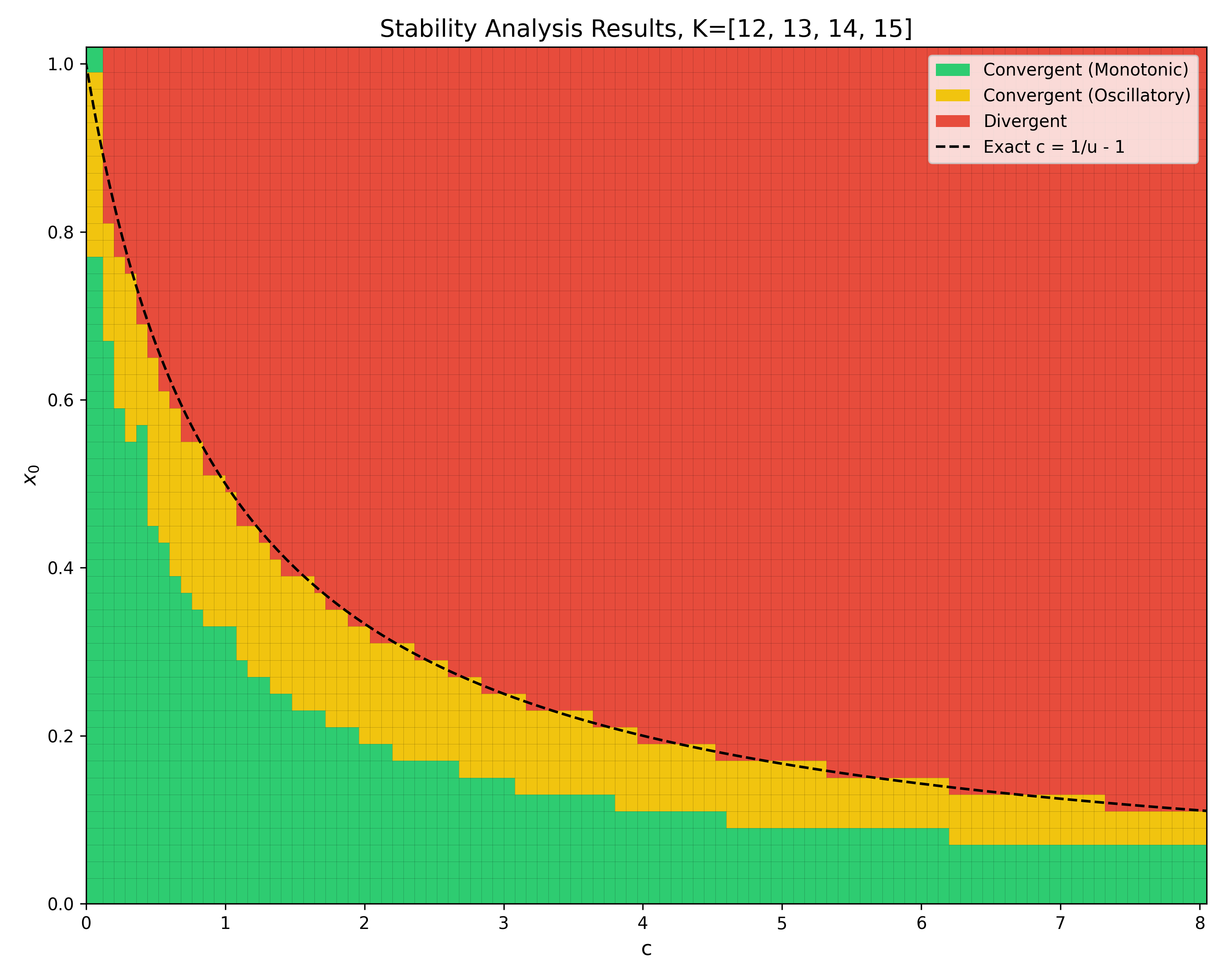}
    \caption{Phase diagram in the $(c,x_0)$ plane for the logistic equation, indicating regimes where the conformally mapped Carleman series converges monotonically (green), oscillatory (yellow), and diverges (red).  This visualization makes explicit how the map parameter $c$ and the initial population $x_0$ jointly determine the success of the regularization.}
    \label{fig:logistic_c-u}
\end{figure}

Before concluding this section, we also present the numerical distribution of $(c, x_0)$ and compare it with the analytical result in Fig.~\ref{fig:logistic_c-u}. Such a comparison is particularly useful when extending the conformal mapping method to other nonlinear differential equations for which exact solutions are not available. In this calculation, the distribution is sampled over $c \in [0,8]$ with a step size of $0.08$, and $x_0 \in [0.02,1.00]$ with a step size of $0.02$. To characterize the convergence properties, we examine the results for truncation parameters $M=K = 12, 13, 14, 15$, and classify the points as follows:
\begin{itemize}
    \item \textbf{Convergent (monotonic):} All final values remain below \(1.005\), corresponding to a \(0.5\%\) tolerance above the strict convergence bound of \(1.00\). The approximations exhibit an ordered behavior with respect to \(K\), where smaller \(K\) yields a smaller convergent value.
    \item \textbf{Convergent (oscillatory):}   All final values remain below \(1.10\), but some exceed \(1.005\) or the sequence of approximations is not ordered in \(K\).
    A representative example is the \(c=8\) panel shown in Fig.~\ref{fig:Logis_conver}.
    \item \textbf{Divergent:} One or more final values exceed \(1.10\), indicating breakdown of the mapped Carleman series.
\end{itemize}

\begin{figure}[H]
    \centering
    \includegraphics[width=0.60\linewidth]{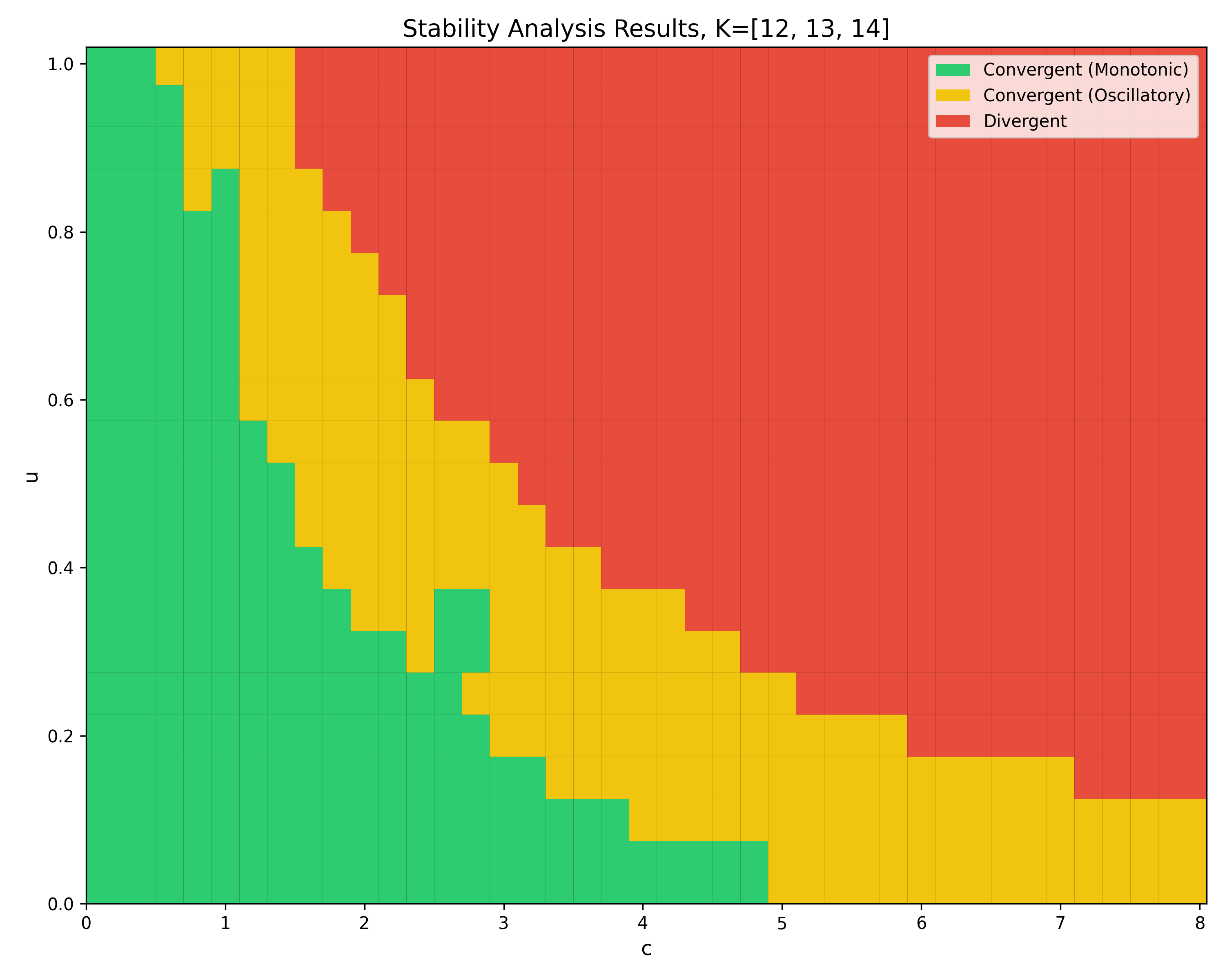}
    \caption{Phase diagram in the $(c,u_2)$ plane for the $L=3$ KPP–Fisher equation, where $u_2$ encodes the initial slope of the reaction–diffusion front.  Monotonically convergent (green), oscillatory convergent (yellow) and divergent (red) regions highlight how the conformal map parameter $c$ controls long‐time behavior in this physical system.}
    \label{fig:KPP_c-u}
\end{figure}

\section{The KPP–Fisher Equations}\label{sec:KPP}
So far, we have shown how to fix the divergence in the Carleman linearization with conformal map and regularized function $f_{M,c}$. Next, we are going to apply it to other diffusion-reaction-type of  nonlinear differential equations whose exact solutions are not available without specific conditions. Here, we consider the KPP–Fisher equation, a typical type of reaction–diffusion equation, given by
\begin{equation}
    \frac{\partial u}{\partial t}=\frac{\partial^2 u}{\partial z^2}+u(1-u)
\end{equation}
where $u(z, t) \in \mathbb{R}$ is the traveling wave solutions. The
first term on the right-hand side is a linear diffusion term (Laplacian term) and the second term is a second-order nonlinear source term (reaction term).

To compute the Laplacian term numerically, we need to discretize it with respect to $z$. Assuming that the computational point $i \in \mathbb{Z}$, equally spaced with width $\Delta z$, was at coordinate $z = i\Delta z $, and applying second-order precision central differencing to the diffusion term. The differential equation then can be approximated by the following difference equation

\begin{equation}\label{eq:KPP_Fisher}
    \frac{d u_i}{d t}=\frac{1}{\Delta z^2}\left(u_{i-1}-2 u_i+u_{i+1}\right)+u_i\left(1-u_i\right),
\end{equation}
where $u_i \in \mathbb{R}$. By setting $\Delta z^2=1$ and imposing the periodic boundary condition $u_i = u_{i+L}$, the KPP–Fisher equation now becomes the finite-dimensional nonlinear differential equation for state $\boldsymbol{u}=\left(u_0, u_1, \cdots, u_{L-1}\right) \in \mathbb{R}^{L}$. Furthermore, due to the diffusion term, there exists negative eigenvalues which are not related to the divergence. Hence, the regularized function should be generalized into
\begin{equation}\label{eq:other_regularized_func}
f_{M,c}(k,t) =
\begin{cases}
I_{1-\omega}(k, M-k+1), & k > 0,\\
1, & k\leq0.
\end{cases}
\end{equation}  
Different spatial length $L$ will lead to different eigenvalue spectrum. Especially for $L=3$, the eigenvalues are all integers and we will first focus on this case and generalize into $L=5$ case which is one of the general case with non-integer eigenvalues.

\begin{figure}
\centering
\includegraphics[width=\linewidth]{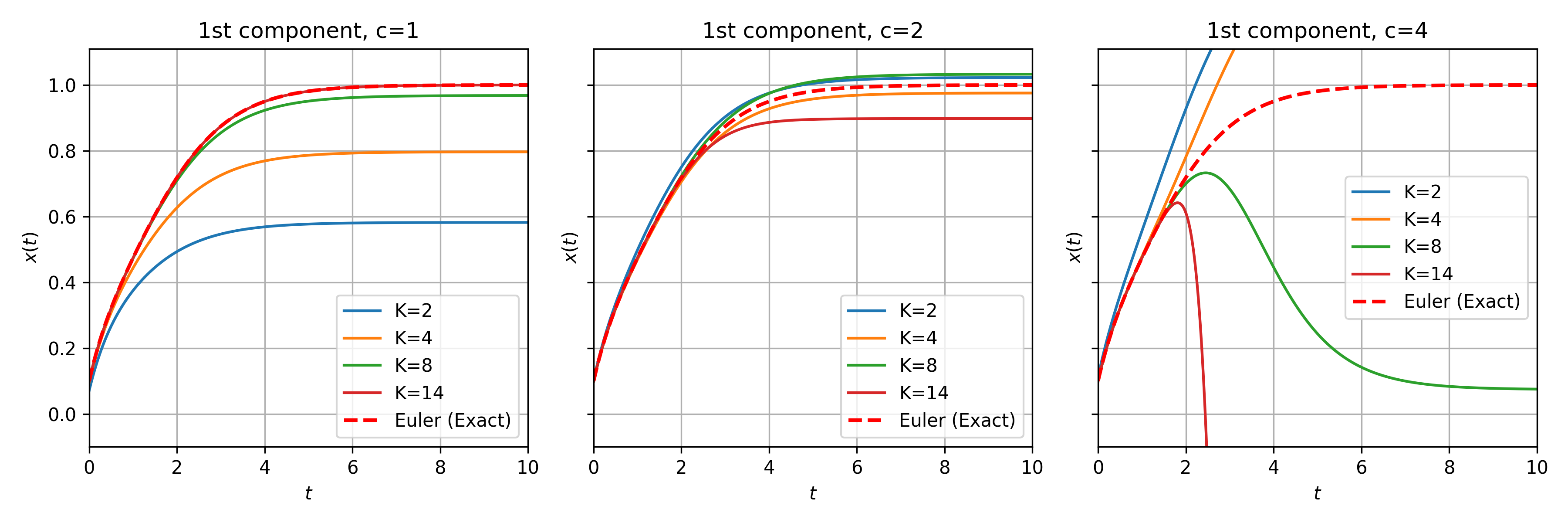}  
\caption{First component of the Carleman solution to the $L=3$ KPP--Fisher equation with truncation up to $M=K=14$ and $s=10^{-4}$}  
\label{fig:KPP_L3_j0}  
\end{figure}

\begin{figure}
\centering
\includegraphics[width=\linewidth]{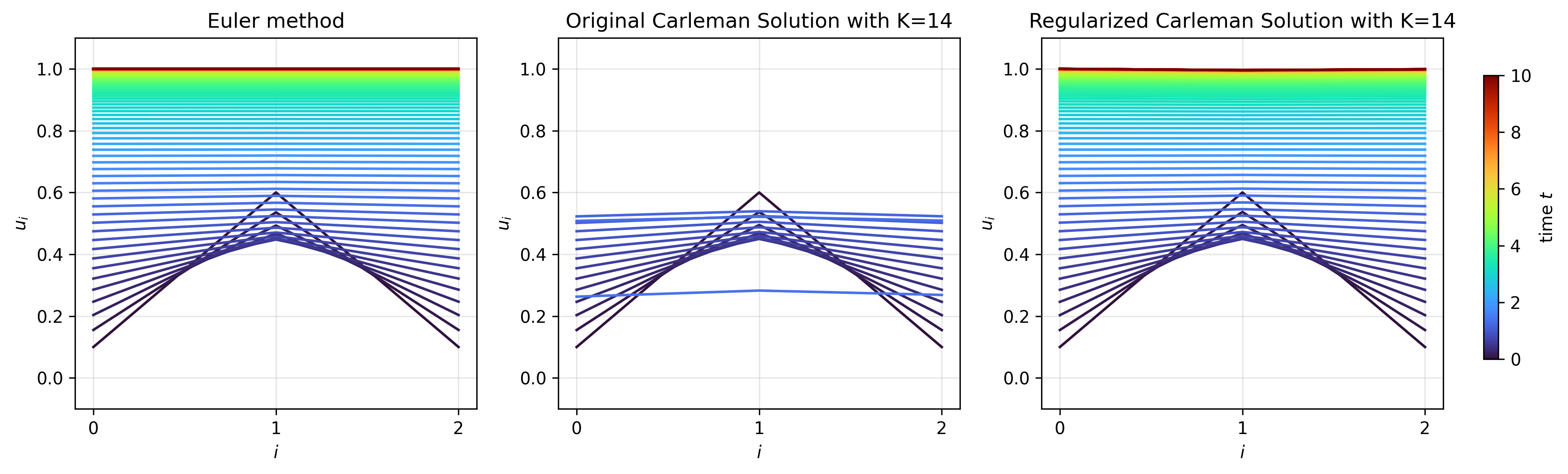}  
\caption{Comparision among the Euler's method, the original Carleman solution and the regularized Carleman solution to the $L=3$ KPP--Fisher equation with truncation $M=K=14$, $c=1$ and $s=10^{-4}$}  
\label{fig:KPP_L3_family}  
\end{figure}

\subsection{The $L=3$ KPP--Fisher Equation}  

In contrast to the logistic equation, there is no general formula for eigenvalues or eigenvectors of the KPP-Fisher systems. For truncation order $K$, the corresponding eigenvalues are integers ranging from $-2K$ to $K$, with a high degree of degeneracy. The general solution therefore takes the form of a Laurent series:  
\begin{equation}
\mathbf{y} = \lim_{K \to \infty} \left( \sum_{k=-2K}^{K} \sum_{i=1}^{g_k} a_{k,i}\, \zeta^k \mathbf{y}^{(k)}_i \right)   
= \lim_{K \to \infty} \left( \sum_{k=-2K}^{K} \mathbf{a}_{k}\, \zeta^k \right),  
\end{equation}  
where $g_k$ denotes the degeneracy of eigenvalue $k$, the eigenvectors satisfy the orthogonality relation  
\[
\mathbf{y}^{(k)}_i \cdot \mathbf{y}^{(l)}_j = \delta_{ij}\delta_{kl},  
\]  
and the coefficient vectors are defined as  
\begin{equation}\label{eq:dege_coef}
\mathbf{a}_{k} = \sum_{i=1}^{g_k} a_{k,i} \mathbf{y}^{(k)}_i.  
\end{equation} 

Applying the conformal map \eqref{eq:conformal_map}, the solution can be expressed as  
\begin{equation}\label{eq:KPP_conformal_map}  
\begin{split}  
\mathbf{y} &= \lim_{K \to \infty} \Bigg(  
\sum_{k=0}^{2K} \mathbf{a}_{-k}\, \zeta^{-k}   
+ \sum_{k=1}^{K} \mathbf{a}_{k}\, \zeta^k I_{1-\omega}(k, M-k+1)\Bigg), 
\end{split}  
\end{equation}  
where the negative-eigenvalue contributions are kept outside the conformal map since they are already convergent.  

The first three components $y_1=u_1,\; y_2=u_2$and $y_3=u_3$ correspond to the solution of the KPP--Fisher equation. To illustrate this, we choose the initial condition $\mathbf{u}(0) = [0.1,\, 0.6,\, 0.1]$, which ensures that $u_1 = u_3$. However, due to the degeneracies in the spectrum, numerical computation of eigenvectors becomes unstable for larger $K$, making accurate approximation with conformal map increasingly difficult.  

To overcome this difficulty, we introduce a small perturbation to the KPP--Fisher equation \eqref{eq:KPP_Fisher}:  
\begin{equation}  
\frac{d u_i}{dt} = \frac{1}{\Delta z^2}\left(u_{i-1} - 2u_i + u_{i+1}\right) + u_i(1 - u_i) + \varepsilon_i u_i,  
\end{equation}  
which modifies the local growth rate from $1$ to $1+\varepsilon_i$ at each site $i$. The perturbation parameters are chosen as  
\begin{equation}  
\varepsilon_i = s \cdot (K+1)^{\, (i-L+1)},  
\end{equation}  
where $s$ is a small random parameter bounded above by $10^{-4}$. This perturbation breaks the degeneracy and allows the eigenvectors to be computed reliably.

Here we employ Euler’s method with time step $\Delta t = 0.01$ to approximate the exact solution. In Carleman linearization, we set $M=K$, initial and final times $t_0=0$, $t_f=10$, $\Delta z=1$, and time step $\Delta t = 0.01$, then the regularized Carleman solutions are shown in Figure~\ref{fig:KPP_L3_j0} and Figure~\ref{fig:KPP_L3_family}. 

Finally, follow the same criteria of convergence and stability in the logistic equation, one can also plot the $c-u_2$ distribution figure, where have fixed $u_0=u_2=0.1$. The distribution is sampled over $c \in [0,8]$ with a step size of $0.20$, and $x_0 \in [0.00,1.00]$ with a step size of $0.05$. To characterize the convergence properties, we examine the results for truncation parameters $K = 12, 13, 14$.

\subsection{The $L=5$ KPP–Fisher Equations}
\begin{figure}
\centering
\includegraphics[width=\linewidth]{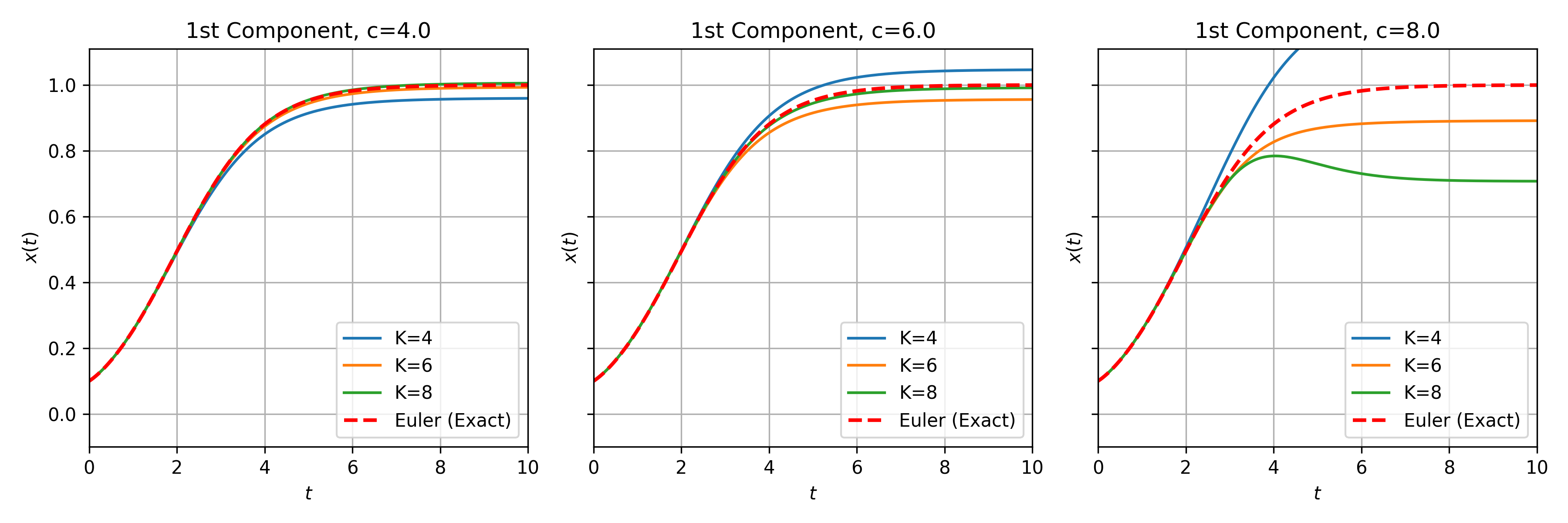}  
\caption{First component of the Carleman solution to the $L=5$ KPP--Fisher equation with truncation up to $M=K=8$ and $s=10^{-4}$}  
\label{fig:KPP_L5_j0}  
\end{figure}

The case $L=3$ is special: all eigenvalues become integers. For general $L$, the eigenvalues $\{k\}$ are real (typically non-integer). 
Fortunately, the regularized function $f_M(k,t)$ can be directly generalized into non-integer $k$ without other modification.
In the $L=5$ case, the first five components of $\mathbf{y}$ correspond to the physical solution of the KPP–Fisher equation. We adopt a symmetric, mild initial condition
\[
\mathbf{u}_0 = [\,0.1,\, 0.1,\, 0.2,\, 0.1,\, 0.1\,],
\]
which enforces $u_1=u_5$ and $u_2=u_4$ by construction. And all remaining numerical settings mirror the $L=3$ case. The regularized Carleman solutions are shown in Figure~\ref{fig:KPP_L5_j0} and Figure~\ref{fig:KPP_L5_family}.

\begin{figure}
\centering
\includegraphics[width=\linewidth]{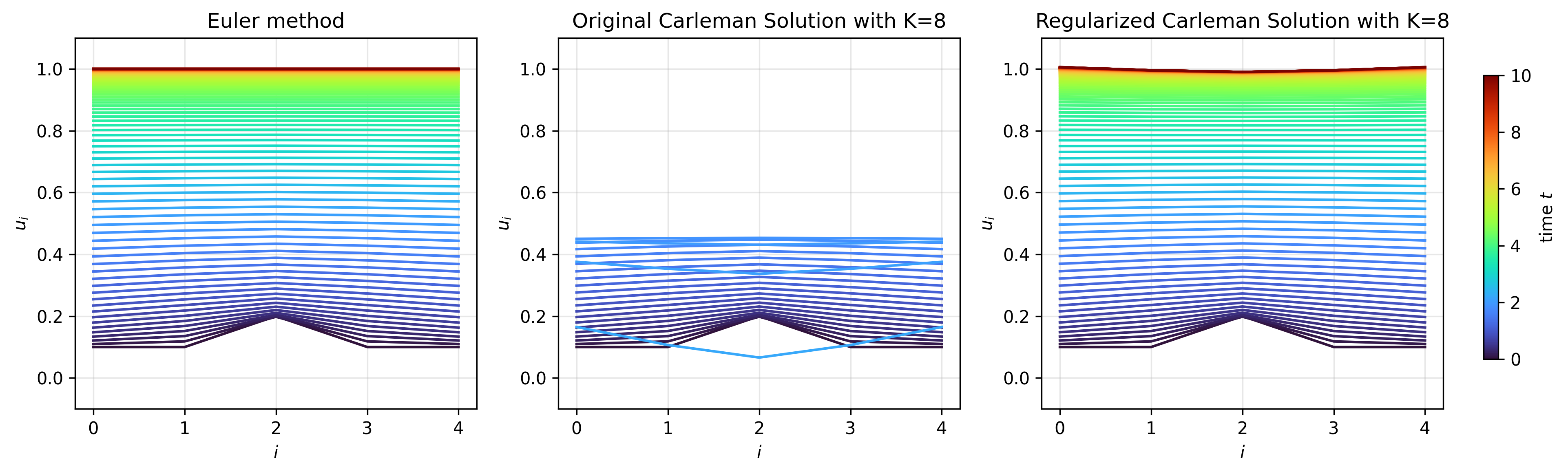}  
\caption{Comparision among the Euler's method, the original Carleman solution and the regularized Carleman solution to the $L=5$ KPP--Fisher equation with truncation $M=K=8$, $c=4$ and $s=10^{-4}$}  
\label{fig:KPP_L5_family}  
\end{figure}

\section{The Phase-Field Model}\label{sec:PFM}

We next turn to discrete phase–field models, which describe interface evolution in materials, and outline how our conformal-mapping technique adapts to cubic nonlinearities.

We have so far examined differential equations with a quadratic nonlinearity of the form 
\(x(1-x)\). The analytical continuation method developed in that setting can be generalized to a
cubic nonlinearity. In this section, we study the cubic logistic equation
\begin{equation}\label{eq:cubic_log}
    \dot{\phi}(t,z) = \phi(1-\phi^{2}), 
    \qquad 0 < \phi_{0} < 1,
\end{equation}
whose solution is known in closed form:
\begin{equation}\label{eq:cubic_log_sol}
    \phi(t)
    = \frac{\phi_{0} e^{t}}{\sqrt{(1-\phi_{0}^{2}) + \phi_{0}^{2} e^{2t}}}.
\end{equation}

Although Carleman linearization may be applied to obtain numerical solutions, the resulting series
diverges for large~\(t\) too, due to the presence of positive eigenvalues. Furthermore, applying the same
conformal mapping \eqref{eq:conformal_map} used for the quadratic case does not resolve the divergence. To understand this behavior, we examine the analytic structure of the solution. Let us still set $\zeta=e^{t}$. The
branch points of Eq.~\eqref{eq:cubic_log_sol} occur at
\begin{equation}
    \zeta_{\pm} = \pm i \frac{\sqrt{1-\phi_{0}^{2}}}{\phi_{0}},
\end{equation}
which lie on the imaginary axis for \(0 < \phi_{0} < 1\). Under the conformal map \eqref{eq:conformal_map}, these singularities are transformed to
\begin{equation}
    \omega_{\pm}
    = \frac{\zeta_{\pm}}{\zeta_{\pm} + c},
    \qquad 
    |\omega_{\pm}|
    = \frac{1}{\sqrt{1 + \left(c/|\zeta_{\pm}|\right)^{2}}}
    < 1
    \quad (\forall\, c > 0).
\end{equation}
Hence, for all positive \(c\), the mapped singularities lie inside the unit disk. Since the
time \(t \ge 0\) corresponds to \(\omega \in [0,1)\), the Carleman series necessarily loses convergence 
before \(\omega\) approaches~1.

To remove this obstruction, we introduce the new variable
\(\eta = e^{2t}\). Expressed in terms of \(\eta\), the
solution becomes
\begin{equation}
    \phi(t)
    = \frac{\phi_{0} \sqrt{\eta}}{\sqrt{(1-\phi_{0}^{2}) + \phi_{0}^{2} \eta}},
\end{equation}
with branch point
\begin{equation}
    \eta_{*}
    = -\frac{1-\phi_{0}^{2}}{\phi_{0}^{2}},
\end{equation}
which now lies on the negative real axis. Under the modified conformal map
\begin{equation}
    \eta = \frac{c\omega'}{1-\omega'},
    \qquad 
    \omega' = \frac{\eta}{\eta + c},
\end{equation}
the branch point lies outside the unit disk whenever 
\(c < (1-\phi_{0}^{2})/\phi_{0}^{2}\). In this case, In this case, the regularized function is modified into
\begin{equation}
f_{M,c}(k,t) =
\begin{cases}
I_{1-\omega}(\frac{k}{2}, M-\frac{k}{2}+1), & k > 0,\\
1, & k\leq0.
\end{cases}
\end{equation}  
and the divergence can be fixed.

\begin{figure}[H]
    \centering
    \includegraphics[width=0.8\linewidth]{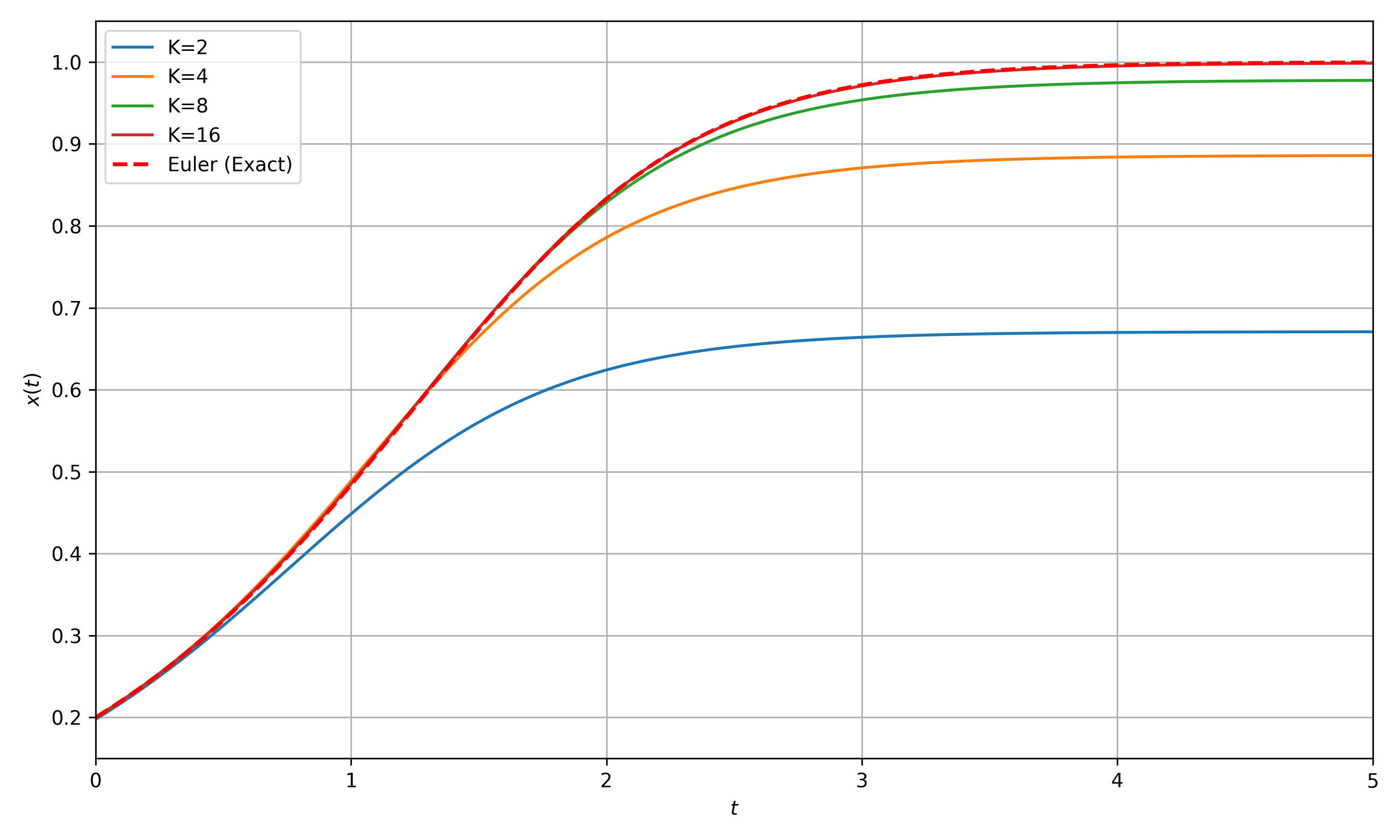}
    \caption{
        Time evolution of the cubic logistic solution after applying the new conformal map,
        for \(x_{0}=0.2\), scale parameter \(c=5\), truncation \(M=K\), and 
        \(K \le 16\).
    }
    \label{fig:cove_phm}
\end{figure}

With this modification, the numerical approximation takes the form
\begin{equation}
\begin{split}
    \mathbf{y}(t)
    &= \lim_{K\to\infty}
        \left(
            \sum_{k=-2K}^{K} \mathbf{a}_{k} e^{k t}
        \right) \\
    &= \lim_{K \to \infty} \Bigg(  
\sum_{k=0}^{2K} \mathbf{a}_{-k}\, \zeta^{-k}   
+ \sum_{k=1}^{K} \mathbf{a}_{k}\, \zeta^k I_{1-\omega'}\left(\frac{k}{2}, M-\frac{k}{2}+1\right)\Bigg).
\end{split}
\end{equation}
where we follow the same notation $\{ a_k\}$ defined in Eq.\eqref{eq:dege_coef}. Figure~\ref{fig:cove_phm} shows the resulting numerical convergence.

\subsection{The Discrete Phase-Field Model}

The cubic logistic equation discussed above can be viewed as a simplified phase–field model with the
diffusion term removed:
\begin{equation}
    \dot{\phi}(t,z)
    = \phi_{zz} + \phi(1-\phi^{2}).
\end{equation}
In phase–field dynamics, the initial condition strongly influences the limiting state: a strictly 
positive (negative) initial condition drives the solution toward \(+1\) (\(-1\)). By analogy with the
KPP--Fisher equation, we discretize the spatial domain to obtain
\begin{equation}
    \frac{d\phi_{i}}{dt}
    = \frac{1}{\Delta z^{2}}
        \left(\phi_{i-1} - 2\phi_{i} + \phi_{i+1}\right)
      + \phi_{i}(1-\phi_{i}^{2}),
    \qquad \phi_{i} \in \mathbb{R}.
\end{equation}
We impose periodic boundary conditions with period \(L\) and choose \(\Delta z = 1\). Also, the regularized function is generalized into Eq.\eqref{eq:other_regularized_func} due to the diffusion term

As before, the standard Carleman linearization diverges for large~\(t\), but the new conformal map 
restores convergence. We illustrate this for the case \(L = 3\), where the eigenvalues are all integers, with truncation
parameters \(M = K \le 14\),
initial conditions \(\boldsymbol{\phi}_{0} = [0.1, 0.4, 0.1]\), scale parameter \(c = \{ 5, 15, 20\}\) for positive initial value case; \(\boldsymbol{\phi}_{0} = [0.1, -0.5, 0.1]\) and \(c = \{ 20, 40, 80\}\) for negative initial value.

\begin{figure}
\centering
\includegraphics[width=\linewidth]{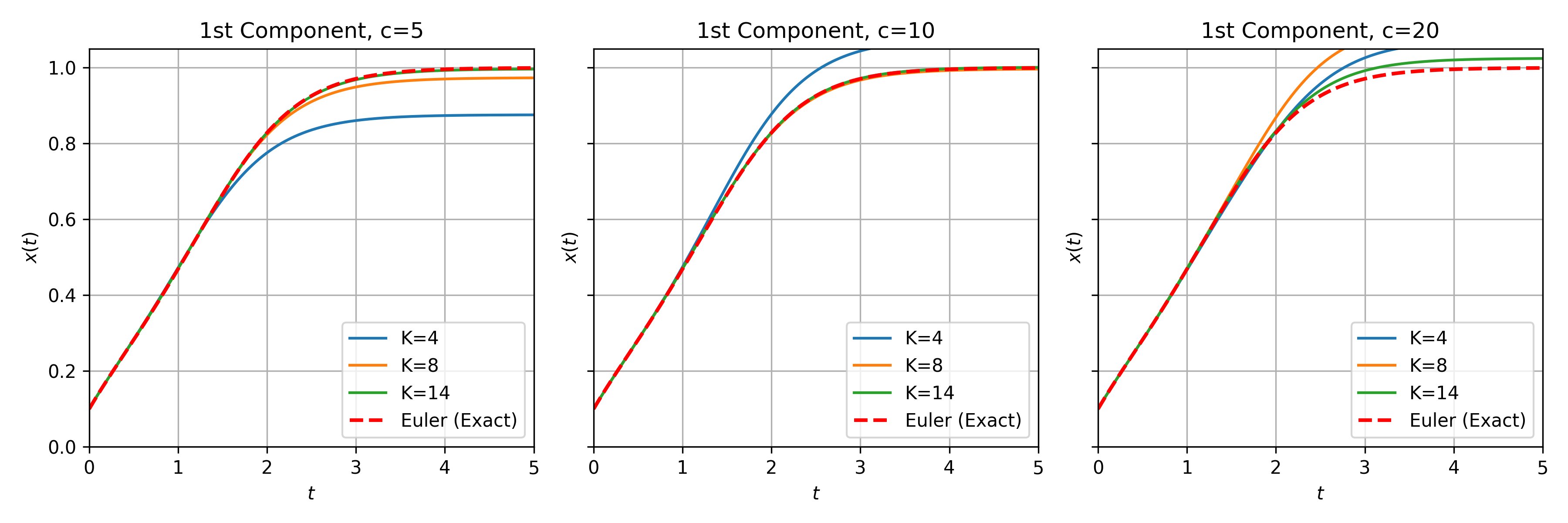}  
\caption{First component of the Carleman solution to the $L=3$ phase–field model with initial value \(\boldsymbol{\phi}_{0} = [0.1, 0.4, 0.1]\), truncation up to $M=K=14$ and $s=10^{-4}$}  
\label{fig:pfm_L3_j0}  
\end{figure}

\begin{figure}
\centering
\includegraphics[width=\linewidth]{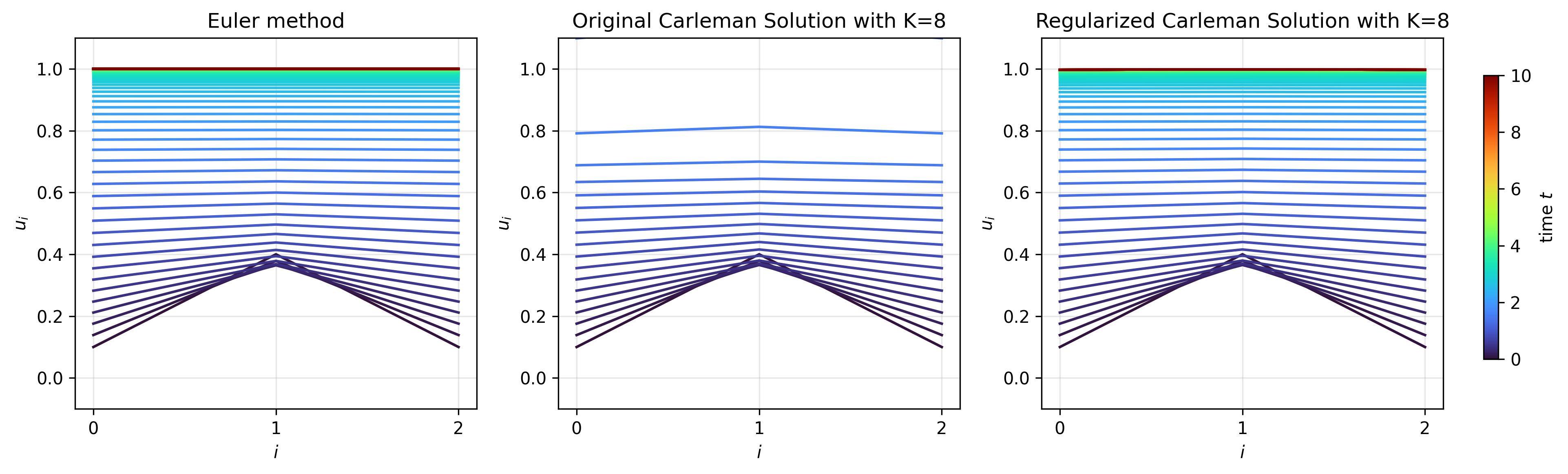}  
\caption{Comparision among the Euler's method, the original Carleman solution and the regularized Carleman solution to the $L=3$ phase–field model with initial value \(\boldsymbol{\phi}_{0} = [0.1, 0.4, 0.1]\), truncation $M=K=14$, $c=5$ and $s=10^{-4}$}  
\label{fig:KPP_L5_family}  
\end{figure}

\begin{figure}
\centering
\includegraphics[width=\linewidth]{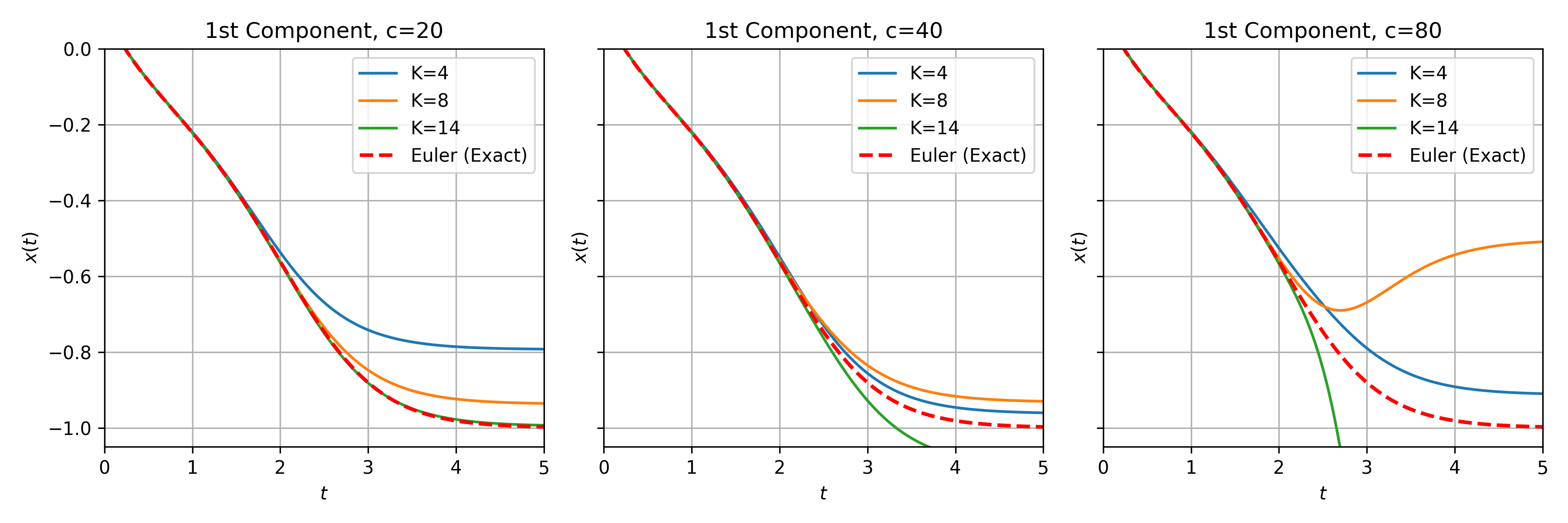} 
\caption{First component of the Carleman solution to the $L=3$ phase–field model with initial value \(\boldsymbol{\phi}_{0} = [0.1, -0.5, 0.1]\), truncation up to $M=K=14$ and $s=10^{-4}$}  
\label{fig:pfm_L3_j0}  
\end{figure}

\begin{figure}
\centering
\includegraphics[width=\linewidth]{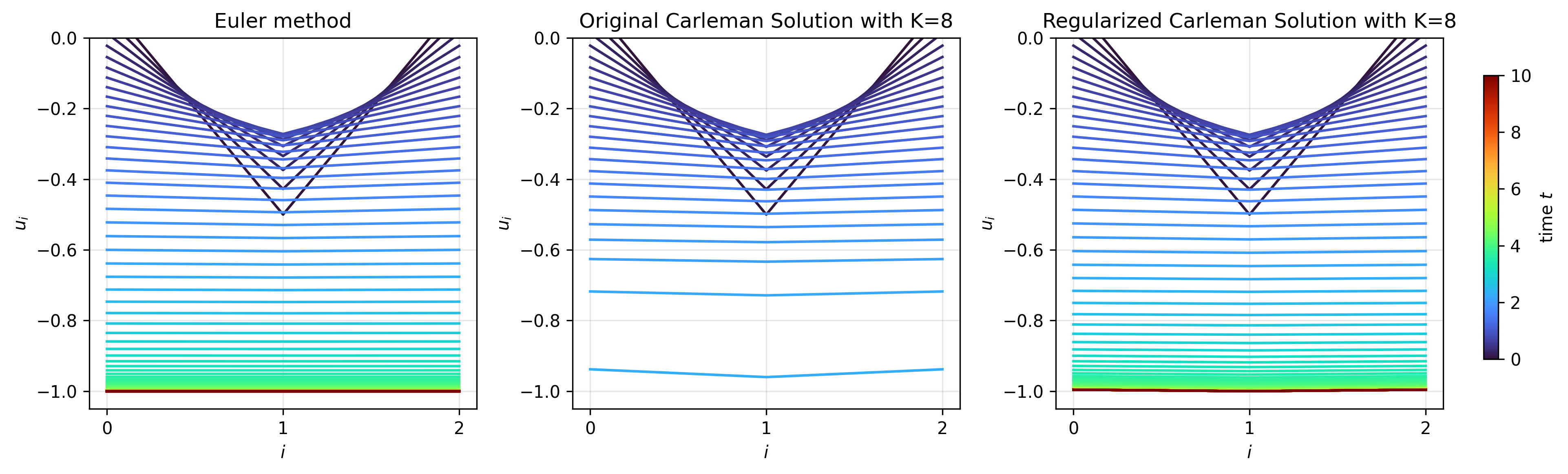}  
\caption{Comparision among the Euler's method, the original Carleman solution and the regularized Carleman solution to the $L=3$ phase–field model with initial value \(\boldsymbol{\phi}_{0} = [0.1, -0.5, 0.1]\), truncation $M=K=14$, $c=5$ and $s=10^{-4}$}  
\label{fig:KPP_L5_family}  
\end{figure}

Finally, we present the $(c,\phi_2)$ stability diagram, which is analogous to the one obtained for the $L=3$ KPP--Fisher model. Since the behavior of the solution is determined by the sign of the average initial value $\boldsymbol{\phi}_{0}$, we separate the results into two cases based on the critical point $\phi_2 = -0.2$ (with $\phi_1 = \phi_3 = 0.1$): $\phi_2 < -0.2$ and $\phi_2 \geq -0.2$. 

The parameter space is sampled over $c \in [0,8]$ (and $[0, 40]$) with a step size of $0.20$ (and $1.0$), while $x_0 \in [0.00,1.00]$ (and $[-0.20,-1.20]$) is sampled with a step size of $0.05$ for the $\phi_2 \geq -0.2$ (and $\phi_2 < -0.2$) case, respectively. To characterize the convergence properties, we examine the results across truncation parameters $K = 10, 11, 12, 13$.

\begin{figure}[htbp]
    \centering
    \begin{subfigure}[b]{0.49\textwidth}
        \centering
        \includegraphics[width=\textwidth]{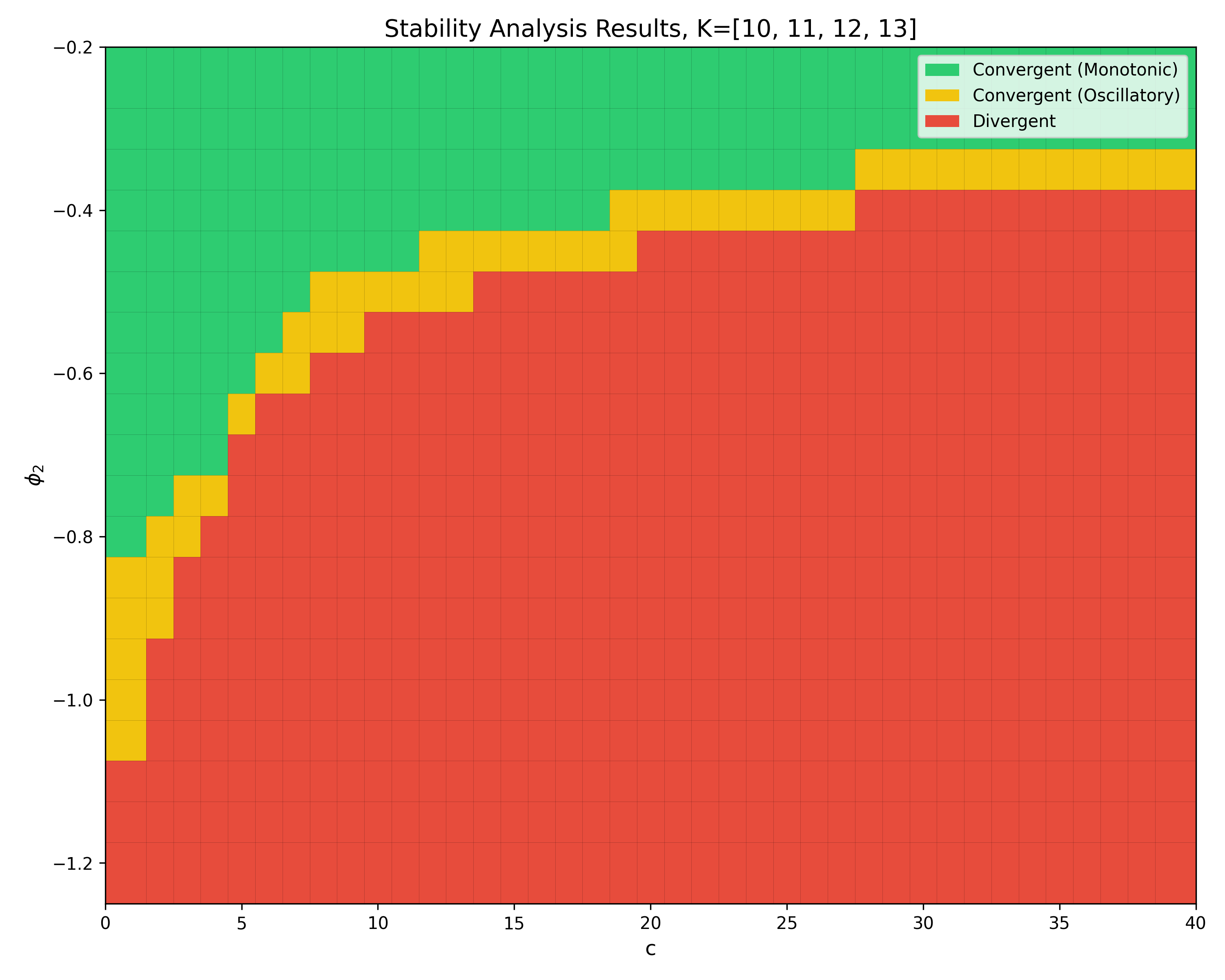}
        \caption{Stability analysis for $\phi_2 < -0.2$.}
        \label{fig:stability_neg}
    \end{subfigure}
    \hfill 
    \begin{subfigure}[b]{0.49\textwidth}
        \centering
        \includegraphics[width=\textwidth]{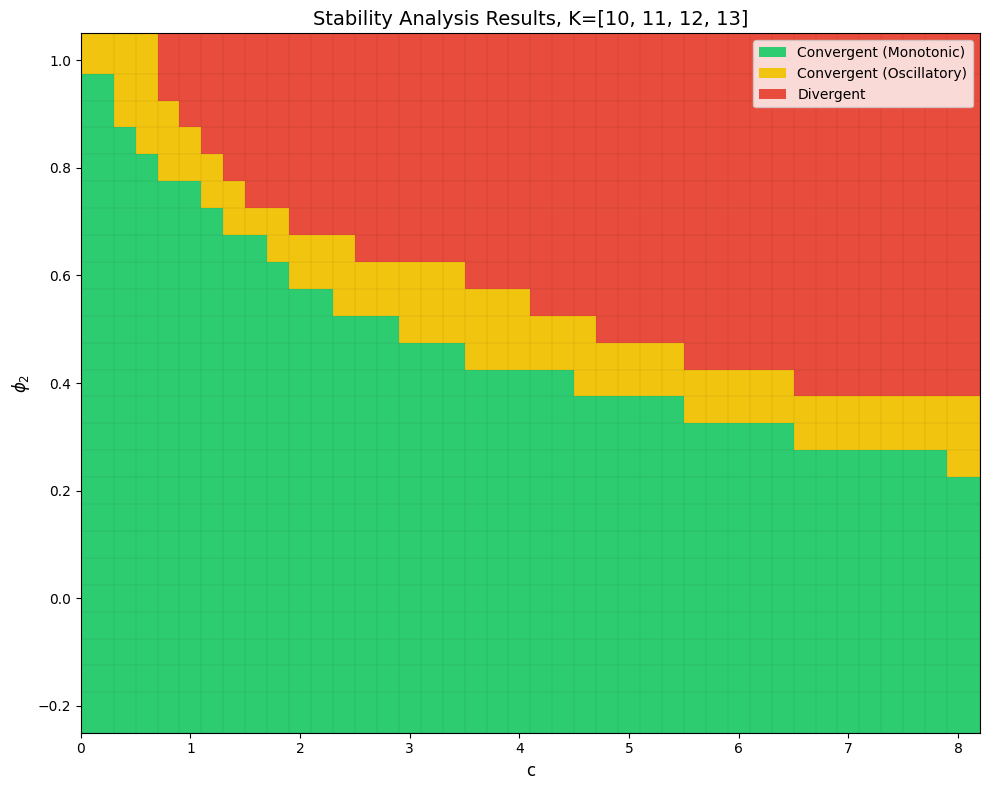}
        \caption{Stability analysis for $\phi_2 \geq -0.2$.}
        \label{fig:stability_pos}
    \end{subfigure}
    
    \caption{Phase diagram in the $(c,\phi_2)$ plane for the $L=3$ phase-field model, where $\phi_2$ encodes the initial slope of the reaction--diffusion front. Monotonically convergent (green), oscillatory convergent (yellow), and divergent (red) regions highlight how the conformal map parameter $c$ controls long-time behavior in this physical system.}
    \label{fig:pfm_c-u}
\end{figure}

\section{Quantum Implementation}\label{sec:QSVT}

We now describe how to implement the regularized Carleman method on a quantum computer. The central idea is to apply an eigenvalue transformation to the Carleman matrix $A$. This enables quantum simulation using techniques such as the Linear Combination of Unitaries (LCU) \cite{Childs:2019hts, Meister:2020qfp} and Quantum Singular Value Transformation (QSVT) \cite{Gilyen:2018khw, Martyn:2021eaf}, provided that the Carleman matrix is Hermitian.
\subsubsection*{Basic Idea}
Recall that the Carleman matrix $A$ admits a diagonal decomposition in terms of its eigenvalues ${k}$,
\begin{equation}
A = P \operatorname{diag}({k}) P^{-1},
\end{equation}
which yields the general (divergent) solution to the Carleman linearization \eqref{eq:gen_sol}:
\begin{equation}
\mathbf{y}(t) = P \operatorname{diag}(\{e^{kt}\}_k) P^{-1} \mathbf{y}(0).
\end{equation}
Here, $P$ denotes the eigenvector matrix of $A$. The same decomposition ${P, P^{-1}}$ can be used to construct the regularized solution \eqref{eq:regularized_sol}, which takes the form
\begin{equation}\label{eq:matrix_regularized_sol}
\mathbf{y}(t) = P \operatorname{diag}(\{e^{kt} f_{M,c}(k,t)\}_k) P^{-1} \mathbf{y}(0).
\end{equation}

We define the corresponding eigenvalue transformation as
\begin{equation}
k \mapsto e^{kt} f_{M,c}(k,t).
\end{equation}

If $A$ is Hermitian, this becomes a singular value transformation, and QSVT can be directly applied to implement the operator
\begin{equation}
 P \operatorname{diag}(\{e^{kt} f_{M,c}(k,t)\}_k) P^{-1},
\end{equation}
thus yielding the regularized solution \eqref{eq:matrix_regularized_sol} at time $t$.

For non-Hermitian $A$, we approximate the transformed function by a polynomial $p_n$ of degree $n$,
\begin{equation}\label{eq:poly_approx}
e^{kt} f_{M,c}(k,t) \approx p_n(k,t)= \sum_{i=0}^{n} a_i(t) k^i,
\end{equation}
which leads to the polynomial approximation
\begin{equation}\label{eq:poly_sol}
\mathbf{y}(t) \approx p_n(A,t)\mathbf{y}(0) =\sum_{i=0}^{n} a_i(t) A^i \mathbf{y}(0).
\end{equation}
This form can be efficiently implemented using the LCU method.

\subsubsection*{Block Encoding}
To apply the idea above, we must address the fact that the Carleman matrix is generally non-unitary (though sparse). We therefore embed it into a larger unitary operator $U(A)$ via block encoding:
\begin{equation}
\frac{A}{s \max_{i,j} |A_{ij}|} = (\bra{0\cdots 0} \otimes I)\, U(A)\, (I \otimes \ket{0\cdots 0}),
\end{equation}
where $s$ denotes the sparsity of $A$, $\max_{i,j} A_{ij}$ is the maximum matrix element, and $I$ is the identity matrix. For further convenience, we define
\begin{equation}\label{eq:BE_factor}
\alpha_{\rm BE}=s\max_{i,j}|A_{ij}|.
\end{equation}
In general, block encoding is composed of three oracles:
\begin{equation}
    U(A) =D_s^{\dagger}O_FO_HD_s,
\end{equation}
where diffusion (or uniform) operator $D_s$ provides $s$ indices for non-zero entries, $O_H$ prepares non-zero the value of non-zero entries and $O_F$ fixs the position of non-zero entries.
Details are provided in Appendix \ref{sec:BE}. Especially for the logistic equation, sparsity $s=2$ with non-zero entries $A_{ii}=i$ and $A_{i,i+1}=-i$ for $i\in \mathbb{Z}^+$, $i\leq K$.

\subsubsection*{LCU}
Due to the scaling factor $\alpha_{BE}$ in the block encoding \eqref{eq:BE_factor}, the coefficients ${a_i(t)}$ should also be rescaled to construct the state preparation operator $\mathcal{P}$:
\begin{equation}
\mathcal{P}\ket{0}_s := \frac{1}{\Lambda} \sum_{i=0}^{n} \sqrt{|\beta_i|} \ket{i}_s,
\end{equation}
where the subindex $s$ denotes the LCU selection register and the normalization factors $\Lambda=\sum_{i=0}^{n}|\beta_i|$ with
\begin{equation}
\beta_i=a_i(t)\alpha_{\rm BE}^i
\end{equation}
whose signs or phases are absorbed into the LCU select operation
Then, the full LCU operator is then given by
\begin{equation}
\mathcal{W} := (\mathcal{P}^\dagger \otimes I)\, \mathcal{S}\, (\mathcal{P} \otimes I),
\end{equation}
where the select operator $\mathcal{S}$ applies powers of $U(A)$:
\begin{equation}
\mathcal{S} = \sum_{i=0}^{n} \ket{i}\bra{i} \otimes U(A)^i,
\end{equation}
with ancilla qubits initialized in $\ket{0\cdots 0}_a$. Finally, measuring the LCU selection register yields the desired approximation of the solution at time $t$.

\begin{figure}[t]
\centering
\includegraphics[width=\linewidth]{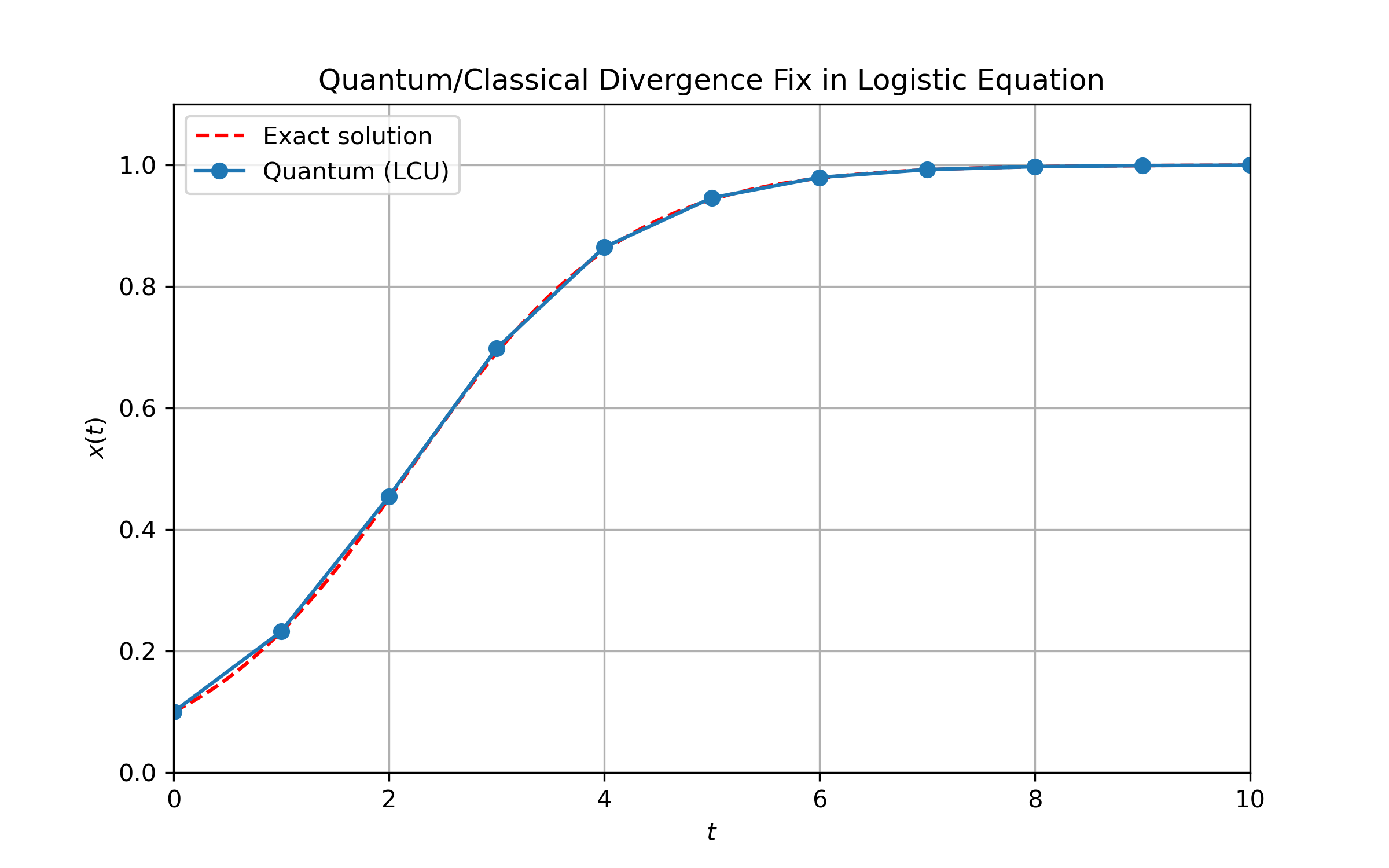}
\caption{Classical simulation of the quantum algorithm applied to the logistic equation with parameters described in the text.}
\label{fig:logistic_eigtransform}
\end{figure}

As a concrete example, we apply this framework to the logistic equation over the time interval $t \in [0,10]$ with time step $\Delta t = 1$, initial condition $x_0 = 0.1$, truncation parameter $K = 3$, regularization parameter $M = 1.5K - 0.5k$, scaling parameter $c = 4$, and interpolation degree $n = K - 1 = 2$. The computation is performed on a classical simulator using the Quri SDK, with measurement omitted. As shown in Fig.~\ref{fig:logistic_eigtransform}, the quantum algorithm provides a good approximation to the exact solution.

Finally, this approach can be generalized to broader classes of nonlinear differential equations, such as diffusion-reaction systems, by extending the regularization function as in Eq.~\eqref{eq:other_regularized_func}. Owing to the continuity of the generalized regularization at $k=0$, polynomial interpolation remains applicable, allowing the same quantum algorithmic framework to be used.

\subsection*{Quantum Complexity and Error Analysis}
We conclude this section with a cost and error analysis of the block-encoding and LCU implementation. We still focus on the logistic equation, for which the conformal-map representation allows an explicit error bound. 
Under the convergence condition \eqref{eq:conver_cond}, the exact logistic solution after the conformal map can be written as
\begin{equation}
x(t) = \frac{r\omega(t)}{1-(1-r)\omega(t)} = r\sum_{m=1}^{\infty}(1-r)^{m-1}\omega(t)^m .
\end{equation}
Therefore, if the mapped series is truncated at order $M$, the regularized Carleman approximation is
\begin{equation}
x_M(t) = r\sum_{m=1}^{M}(1-r)^{m-1}\omega(t)^m ,
\end{equation}
and the map-truncation error is exactly
\begin{equation}
\epsilon_{\rm re}(M,c,t) := |x(t)-x_M(t)| = \frac{r(1-r)^M\omega(t)^{M+1}}{1-(1-r)\omega(t)} .
\end{equation}
For a finite time interval $t\in[0,T]$, this gives the uniform bound
\begin{equation}\label{eq:finite_re_err}
\epsilon_{\rm re}(M,c,T) \leq \frac{r(1-r)^M\omega_T^{M+1}}{1-(1-r)\omega_T}, \qquad \omega_T=\frac{e^T}{e^T+c}.
\end{equation}
In particular, since $\omega(t)<1$ and $\omega(t)\to1$ as $t\to\infty$, the infinite-time bound is
\begin{equation}
\sup_{t\geq0}\epsilon_{\rm re}(M,c,t) \leq (1-r)^M .
\end{equation}
Thus the mapped-series cutoff required to reach regularization error at most $\epsilon_{\rm re}$ may be chosen as
\begin{equation}
M \geq \left\lceil \frac{\log(1/\epsilon_{\rm re})}{-\log(1-r)} \right\rceil .
\end{equation}

We next relate the mapped-series cutoff $M$ to the Carleman truncation order $K$. Because the conformal map satisfies $\phi(0)=0$, the coefficient of $\omega^m$ only depends on Carleman eigenmodes with index $k\leq m$. Consequently, for the logistic equation, if $K\geq M$, then the $K$-truncated Carleman system reproduces the $M$-truncated mapped series exactly. In this case there is no additional Carleman-truncation error beyond the mapped-series error above. This observation motivates the simple choice $K=M$ in the quantum implementation. If $M>K$, additional high-order Carleman modes are omitted, and a separate truncation error must be included.

The remaining error comes from implementing the regularized eigenvalue transformation on the quantum computer. 
Recall that for the non-Hermitian $A$, we use the approximation \eqref{eq:poly_approx}. The polynomial approximation error is given by
\begin{equation}
\epsilon_{\rm poly} = \max_{k\in\sigma(A)} |e^{kt} f_{M,c}(k,t)-p_n(k)|.
\end{equation}
If $A=P_KD_KP_K^{-1}$, then
\begin{equation}
|e^{At} f_{M,c}(A,t)-p_n(A)| \leq \kappa(P_K)\epsilon_{\rm poly},
\end{equation}
where $\kappa(P_K)=\|P_K\| \cdot \|P_K^{-1}\|$
is the eigenvector condition number. This factor is important because the Carleman matrix is generally non-Hermitian and non-normal. For the logistic equation, the eigenvalues are $1,2,\ldots,K$. Therefore, if $p_n$ is chosen as the interpolation polynomial satisfying
\begin{equation}
p_n(k)=e^{kt} f_{M,c}(k,t), \qquad k=1,\ldots,K,
\end{equation}
then $n=K-1$ is sufficient and $\epsilon_{\rm poly}=0$ in exact arithmetic. In practice, finite-precision arithmetic and gate synthesis introduce additional errors.

Next, let $\delta_{\rm BE}$ denote the operator-norm error of one block-encoding query, including arithmetic and rotation-synthesis error. Since the LCU circuit uses powers up to degree $n$, the induced implementation error can be bounded as
\begin{equation}
\epsilon_{\rm BE} = O(\Lambda n\delta_{\rm BE}).
\end{equation}
Thus, to make this contribution at most $\epsilon_{\rm BE}$, it suffices to choose $\delta_{\rm BE} = O\left(\frac{\epsilon_{\rm BE}}{\Lambda n}\right).$

Combining the above contributions, the total error $\epsilon_{tot}(t)=|x(t)-\widetilde{x}(t)|$ in the first component of the logistic solution satisfies
\begin{equation}
\epsilon_{tot}(t)\leq \epsilon_{\rm re}(M,c,t) + \kappa(P_K)\epsilon_{\rm poly} + O(\Lambda n\delta_{\rm BE}) + \epsilon_{\rm meas} + \epsilon_{\rm prep},
\end{equation}
where $\epsilon_{\rm meas}$ is the measurement error and $\epsilon_{\rm prep}$ is the initial-state preparation error. In the parameter regime $K=M$ and $n=K-1$, with exact spectral interpolation, $\epsilon_{re}$ can be replaced by Eq.\eqref{eq:finite_re_err} and $\epsilon_{poly}=0$

We now estimate the gate cost. The block encoding of the logistic Carleman matrix uses $O(\log K)$ system qubits and $O(\log K)$ ancilla qubits. The value oracle contains $O(K)$ controlled rotations, each controlled on $O(\log K)$ qubits. The position oracle is implemented by a controlled constant adder, which also costs $O(\log K)$ Toffoli gates. Hence the Toffoli cost of one block-encoding query is
\begin{equation}
T_{\rm BE}=O(K\log K),
\end{equation}
up to polylogarithmic factors in the inverse gate-synthesis precision.

The LCU circuit uses polynomial degree $n$, and its success probability is governed by the normalization factor $\Lambda$. With oblivious amplitude amplification, the cost of implementing $p_n(A_K)$ scales as
\begin{equation}
T_{\rm LCU} = O(\Lambda n T_{\rm BE}).
\end{equation}
If amplitude estimation is used to estimate the final observable to additive precision $\epsilon_{\rm meas}$, the total Toffoli complexity becomes
\begin{equation}
T_{\rm total} = O\left( \frac{\Lambda n T_{\rm BE}}{\epsilon_{\rm meas}} \right).
\end{equation}
For the logistic implementation with $n=K-1$, this gives
\begin{equation}
T_{\rm total} = O\left( \frac{\Lambda K^2\log K}{\epsilon_{\rm meas}} \right).
\end{equation}
The dependence on $\Lambda$ is important: although increasing $K$ reduces the conformal-map truncation error exponentially as $(1-r)^K$, the LCU normalization may increase with $K$ and with the interpolation coefficients. Therefore, an efficient quantum implementation requires a balance between the regularization parameters $(c,M)$, the Carleman order $K$, the interpolation degree $n$, and the LCU normalization $\Lambda$.

For general reaction--diffusion systems, the same error decomposition applies, but the explicit bound $(1-r)^K$ is special to the logistic equation. In the PDE examples, one must additionally include spatial-discretization error, perturbation error from degeneracy breaking, and the conditioning of the corresponding Carleman eigenbasis. Deriving sharp a priori bounds for these general systems is left as future work.

\section{Conclusion and Discussion}
\label{sec:conclusion}

We investigated the long-time divergence of Carleman linearization for nonlinear differential equations and showed that this divergence is caused by the finite convergence domain of the lifted spectral expansion.  In the logistic equation, the Carleman solution can be written in an eigenbasis as a series in $\zeta=e^t$.  Although this series agrees with the exact solution inside its convergence radius, it diverges along the positive time direction once $\zeta$ leaves that domain.  This gives a direct explanation of the exponential divergence observed in truncated Carleman simulations.

To remove this divergence, we introduced an analytic-continuation method based on a conformal map.  For quadratic nonlinearities such as the logistic and KPP--Fisher equations, the M\"obius map $\zeta=c\omega/(1-\omega)$ sends the positive real time direction into the unit disk in the $\omega$-plane.  In the logistic case, this leads to the explicit convergence condition $c<(1-x_0)/x_0$, which is consistent with the numerical stability diagrams.  We further reformulated the conformal-map correction as an eigenvalue regularization by inserting the regularized function $f_{M,c}(k,t)$ into the divergent Carleman spectral solution.  For positive integer eigenvalues, this function is expressed as the regularized incomplete beta function $I_{1-\omega(t)}(k,M-k+1)$.  Thus, the divergent factor $e^{kt}$ is replaced by the regularized factor $e^{kt}f_{M,c}(k,t)$, while nonpositive eigenvalue contributions are kept unchanged in reaction--diffusion systems.  This form is useful because it turns the analytic-continuation procedure into a direct eigenvalue transformation of the Carleman matrix.

We validated the regularized Carleman method on the logistic equation and on spatially discretized KPP--Fisher equations with periodic boundary conditions.  For the $L=3$ KPP--Fisher system, the truncated Carleman spectrum is integer-valued and the lifted solution takes the form of a Laurent series.  For the $L=5$ case, the spectrum contains noninteger eigenvalues, but the gamma-function definition of $f_{M,c}(k,t)$ applies without essential modification.  In both cases, the regularized Carleman solution suppresses the long-time divergence of the original Carleman approximation and agrees well with the reference solution obtained by direct numerical integration.  The stability diagrams also clarify how the conformal-map parameter $c$ and the initial condition jointly determine the monotonic, oscillatory, and divergent regimes.

For cubic nonlinearities, such as the cubic logistic equation and the phase-field model, a different conformal map is required.  The simple map in $\zeta=e^t$ does not remove the relevant singularities because the branch points of the cubic logistic solution are mapped inside the unit disk.  We therefore introduced the variable $\eta=e^{2t}$ and applied the conformal map $\eta=c\omega'/(1-\omega')$.  This map moves the relevant branch point to the negative real direction in the $\eta$-plane and restores convergence under the corresponding constraint on $c$.  The resulting regularization is obtained by replacing $k$ with $k/2$ in the incomplete-beta-function form of $f_{M,c}(k,t)$ for positive eigenvalues.  This distinction is essential for the phase-field model.  Accordingly, the stability diagram for the discrete phase-field system should be described in the $(c,\phi_2)$ plane, where $\phi_2$ denotes the relevant initial phase-field component, rather than using the KPP--Fisher notation $u_2$.

The role of the mapped-series cutoff $M$ is straightforward.  The Carleman order $K$ determines how many lifted modes are retained, while $M$ determines how far the conformally mapped series is truncated.  In practice, $M$ should be chosen consistently with $K$.  The simple choice $M=K$ already gives stable results in the quadratic examples, while moderately larger values of $M$ can sometimes improve convergence when the Carleman truncation is sufficiently large.  However, taking $M$ too large compared with $K$ may introduce finite-truncation instability.  We therefore regard $M$ as a cutoff coupled to the Carleman order, rather than as an independent parameter requiring extensive optimization.

We also gave a concrete quantum implementation of the regularized Carleman method for the logistic equation.  Since the regularized solution is an eigenvalue transformation of the Carleman matrix, it can be approximated by a polynomial in the Carleman matrix.  For non-Hermitian Carleman matrices, this polynomial form can be implemented by the Linear Combination of Unitaries method after constructing a block encoding of the sparse Carleman matrix.  In the logistic case, the Carleman matrix has a simple sparse structure with two nonzero entries per row, and we explicitly constructed its block encoding using a diffusion operator, a value oracle, and a position oracle.  Combining this block encoding with LCU gives a concrete quantum procedure for preparing the regularized Carleman solution.

The resource estimate shows that the total error consists of the mapped-series truncation error, polynomial approximation error, block-encoding error, measurement error, and state-preparation error.  For the logistic equation, the mapped-series truncation error can be bounded explicitly and decays exponentially with $M$ under the convergence condition.  When $K=M$ and the interpolation polynomial exactly matches the transformed eigenvalues on the truncated spectrum, the polynomial approximation error vanishes in exact arithmetic.  The resulting LCU complexity depends on the block-encoding cost, the polynomial degree, the measurement precision, and the LCU normalization factor.  This analysis shows that the regularized Carleman method is not only a classical divergence-removal technique, but also admits a concrete quantum implementation with an explicit resource estimate.

Several issues remain open.  For general nonlinear differential equations, the optimal conformal map should depend on the singularity structure of the solution, and systematic map selection remains an important problem.  For reaction--diffusion systems, a complete error analysis should also include spatial discretization error, perturbation error from degeneracy breaking, and conditioning effects from the Carleman eigenbasis.  Another important direction is the application of the present regularization strategy to computational fluid dynamics.  Carleman-based approaches have recently been studied in the context of lattice Boltzmann and fluid-dynamical simulations, where long-time divergence of the Carleman solution has also been observed \cite{fluids7010024,Itani:2023ltd,Sanavio:2024tnw}.  Since this divergence has the same qualitative origin as the examples studied in this work, it would be important to test whether the conformal-map regularization and the associated function $f_{M,c}(k,t)$ can stabilize Carleman simulations of nonlinear CFD models.  From the quantum-algorithmic viewpoint, the most important future direction is to develop efficient methods for constructing block encodings of Carleman matrices for general classes of differential equations.  The logistic equation admits a simple explicit construction, but general polynomial systems, reaction--diffusion equations, phase-field models, and fluid-dynamical equations lead to more complicated sparse tensor structures.  An efficient and general block-encoding construction would make the LCU-based regularized Carleman framework applicable to a much wider class of nonlinear dynamics.

In summary, we identified the divergence of Carleman linearization as a convergence-domain problem of the original spectral series and removed it by analytic continuation implemented through the regularized function $f_{M,c}(k,t)$.  The method applies to quadratic and cubic nonlinearities, extends to reaction--diffusion and phase-field models under periodic discretization, and admits a concrete LCU-based quantum implementation with resource estimates.  These results provide a foundation for long-time stable Carleman-based simulation of nonlinear differential equations on both classical and quantum computers.

\section*{Acknowledge}
The authors are partial supported from the New Energy and Industrial Technology Development Organization (NEDO) through the NEDO Challenge program.

\appendix
\section{Block Encoding of Logistic Carleman Matrix}\label{sec:BE}
The logistic Carleman matrix is a sparse matrix with sparsity $s=2$. Let $|*\rangle_{\mathrm{ind}}$ be a computational basis state that takes the row and column indices of $A$ as its arguments. Here, the block encoding is built following by \cite{Camps:2022jnx}

First, we need a diffusion operator $D_s$ to provide $s$ indices for non-zero entries
\begin{equation}
    D_s|0\rangle_{\mathrm{ind}} = \frac{1}{\sqrt{s}} \sum_{i=0}^{s-1}|i\rangle_{\mathrm{ind}}
\end{equation}
Besides, there are two important oracles: $O_H$ and $O_F$.

\subsubsection*{$O_H$ (value)}
    An operation that encodes $A_{xy}$ depending on the row and column indices $(x,y)$:
    \begin{equation}
    O_H \, |\psi\rangle|y\rangle_{\mathrm{ind}} |i\rangle_{\mathrm{ind}} 
    = U(A_{xy_i}) |\psi\rangle|y\rangle_{\mathrm{ind}} |i\rangle_{\mathrm{ind}} .
    \end{equation}
    Here we add one ancilla qubit $|\psi\rangle$ to realize the rotation $R_Y(\theta_{x,y})$ with $\theta_{x,y}$
    \begin{equation}
            \theta_{x,y} = 2\arccos \left(\frac{A_{xy}}{\max_{i,j} A_{ij}}\right)
    \end{equation}
    and
    \begin{equation}
            \begin{aligned}
    R_Y(\theta_{x,y})|0\rangle
    &= \cos\!\left(\frac{\theta_{x,y}}{2}\right)|0\rangle
     + \sin\!\left(\frac{\theta_{x,y}}{2}\right)|1\rangle \\
    &= \frac{A_{xy}}{\max_{i,j} A_{ij}}\,|0\rangle
     + \sqrt{1-\left(\frac{A_{xy}}{\max_{i,j} A_{ij}}\right)^2}\,|1\rangle
    \end{aligned}
    \end{equation}
In the logistic equation, there are only two rotation angeles: $\theta_{diag}$ and $\theta_{off-diag}$
\begin{equation}
\theta_{diag}(y)=2 \arccos \left(\frac{y}{K}\right),\quad \theta_{off-diag}(y)=2 \arccos \left(-\frac{y-1}{K}\right),
\end{equation}
where the column index $y$ for $\theta_{diag}$ should be counted from 2. The value oracle can be realized by controlled-$R_Y$ gates as
\scriptsize
\begin{equation}
\begin{aligned}
    &O_H:\quad \quad \sum_{y=1}^{3}\sum_{i=0}^{1}\frac{1}{\sqrt{2}}\ket{0}|y\rangle_{\text{ind}}|i\rangle_{\text{ind}} \rightarrow  \\
    &\sum_{y=1}^{3}\frac{1}{\sqrt{2}}R_Y(\theta_{diag}(y))\ket{0}|y\rangle_{\text{ind}}|0\rangle_{\text{ind}}+  \sum_{y=1}^{3}\frac{1}{\sqrt{2}}R_Y(\theta_{off-diag}(y))\ket{0}|y\rangle_{\text{ind}}|1\rangle_{\text{ind}}
\end{aligned}
\end{equation}
\normalsize

\subsubsection*{$O_F$ (position)}
    Let $y_i$ denote the column index of the $i$-th nonzero element in row $x$. Then
    \begin{equation}
    O_F \, |y\rangle_{\mathrm{ind}} |i\rangle_{\mathrm{ind}}
    = |x(y,i)\rangle_{\mathrm{ind}} |i\rangle_{\mathrm{ind}} .
    \end{equation}

Combine all of these three operators $O_FO_HD_s$ on $|0\rangle\left|0\right\rangle_{\mathrm{ind}}|x\rangle$, one can see
\scriptsize
\begin{equation}
\begin{aligned}
|0\rangle\left|0\right\rangle_{\mathrm{ind}}|y\rangle & \xrightarrow{D_s} \frac{1}{\sqrt{s}} \sum_{i \in s}|0\rangle|i\rangle|y\rangle \\
& \xrightarrow{O_H} \frac{1}{\sqrt{s}} \sum_{i \in s}\left(\frac{A_{xy(x,i)}}{\max_{i,j} A_{ij}}\,|0\rangle
     + \sqrt{1-\left(\frac{A_{xy}}{\max_{i,j} A_{ij}}\right)^2}\,|1\rangle\right)|i\rangle|y\rangle \\
& \xrightarrow{O_F} \frac{1}{\sqrt{s}} \sum_{i \in s}\left(\frac{A_{xy}}{\max_{i,j} A_{ij}}\,|0\rangle
     + \sqrt{1-\left(\frac{A_{xy}}{\max_{i,j} A_{ij}}\right)^2}\,|1\rangle\right)|i\rangle|x(y, i)\rangle,
\end{aligned}
\end{equation}
\normalsize
on the other hand, 
\begin{equation}
    \bra{x'}\bra{0}_{\mathrm{ind}}\bra{0}D_s^{\dagger} = \frac{1}{\sqrt{s}}\sum_{i'\in s}\bra{x'}\bra{i'}
\end{equation}
Finally, $U(A) =D_s^{\dagger}O_FO_HD_s $ leads to
\begin{equation}
    \bra{x}\bra{0}_{\mathrm{ind}}\bra{0} U(A)|0\rangle\left|0\right\rangle_{\mathrm{ind}}|y\rangle = \frac{A_{xy}}{s\max_{i,j} A_{ij}}
\end{equation}
Especially in the logistic equation, the $O_F$ is realized by an constant adder:
\begin{equation}
O_F: \quad\sum_{i=0}^{1}\frac{1}{\sqrt{2}}|x\rangle|i\rangle_{\text{ind}} \mapsto  \frac{1}{\sqrt{2}}(|x\rangle|0\rangle_{\text{ind}}+|x-1\rangle|1\rangle_{\text{ind}} ,
\end{equation}

Here is one simple example for $K=3$, the exact Carleman matrix is
\begin{equation}
    A=\begin{pmatrix}
      1 & -1 & 0\\
      0 & 2 & -2 \\
      0 & 0 & 3
    \end{pmatrix}
\end{equation}
and the result is embed into a $4\times4$ matrix after block encoding.
\scriptsize
\begin{verbatim}
[[ 1.0000000e+00+0.j -1.0000000e+00+0.j  0.0000000e+00+0.j  0.0000000e+00+0.j]
 [ 0.0000000e+00+0.j  2.0000000e+00+0.j -2.0000000e+00+0.j  0.0000000e+00+0.j]
 [ 0.0000000e+00+0.j  0.0000000e+00+0.j  3.0000000e+00+0.j  1.8369702e-16+0.j]
 [-5.5109106e-16+0.j  0.0000000e+00+0.j  0.0000000e+00+0.j  1.8369702e-16+0.j]]
\end{verbatim}
\normalsize

\bibliographystyle{utphys}
\bibliography{refs.bib}

\end{document}